\documentclass[10pt,journal,compsoc]{IEEEtran}
\ifCLASSOPTIONcompsoc
  \usepackage[nocompress]{cite}
\else
  \usepackage{cite}
\fi
\ifCLASSINFOpdf
  \usepackage[pdftex]{graphicx}
\else
  \usepackage[dvips]{graphicx}
\fi
\graphicspath{{./Figs/}{./figs/}}
\DeclareGraphicsExtensions{.pdf,.jpeg,.png,.eps}
\usepackage{amsmath}
\usepackage{amssymb}
\usepackage{amsthm}
\usepackage{mathtools}
\allowdisplaybreaks
\usepackage{xcolor}
\definecolor{teal}{HTML}{008080}
\usepackage{hyperref}
\hypersetup{
  colorlinks=true,
  citecolor=teal
}

\usepackage[ruled, linesnumbered, noend]{algorithm2e}

\usepackage{array}

\ifCLASSOPTIONcompsoc
 \usepackage[caption=false,font=normalsize,labelfont=sf,textfont=sf]{subfig}
\else
 \usepackage[caption=false,font=footnotesize]{subfig}
\fi

\usepackage{stfloats}
\usepackage{url}

\usepackage{bm}
\usepackage{color}
\usepackage{tikz}
\usetikzlibrary{arrows.meta,positioning}
\usepackage[justification=centering]{caption}
\usepackage{multirow}
\usepackage{blindtext}
\usepackage{adjustbox}
\usepackage{setspace}
\usepackage{xfrac}

\newcommand{\descr}[1]{\vspace{0.2cm} \noindent \textbf{\sffamily #1}}

\newtheorem{definition}{Definition}
\newtheorem{lemma}{Lemma}
\newtheorem{theorem}{Theorem}
\newtheorem{corollary}{Corollary}
\newtheorem{proposition}{Proposition}

\DeclareMathOperator*{\argmax}{arg\,max}

\newcommand{\revise}[1]{{\color{black}#1}}
\newcommand{\reviseA}[1]{{\color{black}#1}}

\begin{document}
%
\title{A Differentially Private Framework for Deep Learning with \revise{Convexified} Loss Functions}
%
%
%

\author{Zhigang~Lu,
        ~Hassan~Jameel~Asghar,
        ~Mohamed~Ali~Kaafar,
        ~Darren~Webb,
        and~Peter~Dickinson
\IEEEcompsocitemizethanks{
\IEEEcompsocthanksitem Z. Lu, H. J. Asghar and M. A. Kaafar are with Department of Computing, Macquarie University, Sydney, Australia.\protect\\
E-mail: \{zhigang.lu, hassan.asghar, dali.kaafar\}@mq.edu.au
\IEEEcompsocthanksitem D. Webb and P. Dickinson are with Cyber \& Electronic Warfare Division, Defence Science \& Technology Group, Edinburgh, Australia.\protect\\
E-mail: \{darren.webb,peter.dickinson\}@dst.defence.gov.au
}
\thanks{Manuscript received Month DD, YYYY; revised Month DD, YYYY.}}

%
%

\markboth{IEEE Transactions on ABCD,~Vol.~xx, No.~xx, Month~YYYY}%
{Lu \MakeLowercase{\textit{et al.}}: A Differentially Private Framework for Deep Learning with \revise{Convexified} Loss Functions}
%



\IEEEtitleabstractindextext{%
\begin{abstract}
Differential privacy (DP) has been applied in deep learning for preserving privacy of the underlying training sets. Existing DP practice falls into three categories~\cite{JayaramanB2019} -- objective perturbation (injecting DP noise into the objective function), gradient perturbation (injecting DP noise into the process of gradient descent) and output perturbation (injecting DP noise into the trained neural networks, scaled by the global sensitivity of the trained model parameters). They suffer from three main problems. First, conditions on objective functions limit objective perturbation in general deep learning tasks. Second, gradient perturbation does not achieve a satisfactory privacy-utility trade-off due to over-injected noise in each epoch. Third, high utility of the output perturbation method is not guaranteed because of the loose upper bound on the global sensitivity of the trained model parameters as the noise scale parameter. To address these problems, we analyse a tighter upper bound on the global sensitivity of the model parameters. Under a black-box setting, based on this global sensitivity, to control the overall noise injection, we propose a novel output perturbation framework by injecting DP noise into a randomly sampled neuron (via the exponential mechanism) at the output layer of a baseline non-private neural network trained with a \revise{convexified} loss function. We empirically compare the privacy-utility trade-off, measured by accuracy loss to baseline non-private models and the privacy leakage against black-box membership inference (MI) attacks~\cite{ShokriR2017}, between our framework and the open-source differentially private stochastic gradient descent (DP-SGD) approaches on six commonly used real-world datasets. The experimental evaluations show that, when the baseline models have observable privacy leakage under MI attacks, our framework achieves a better privacy-utility trade-off than existing DP-SGD implementations, \reviseA{given an overall privacy budget $\epsilon \leq 1$ for a large number of queries.}
\end{abstract}
}
%
%
\maketitle

\IEEEdisplaynontitleabstractindextext

%
\IEEEpeerreviewmaketitle

\IEEEraisesectionheading{
\section{Introduction}
\label{sec:introduction}}

%
%
%
%
\IEEEPARstart{M}{embership} Inference (MI) attacks pose serious privacy risks to Machine Learning-as-a-Service (MLaaS). In a nutshell, MI attacks apply a binary classifier on the prediction vectors (outputs) of data samples obtained from a non-private machine learning (ML) model to infer whether the data samples are members of the training set or not. In this paper, we assume the baseline ML model is a deep neural network (DNN). There are two general types of MI attacks, black-box MI attacks~\cite{ShokriR2017} and white-box MI attacks~\cite{NasrM2018C}, depending on the adversary's access to the target ML model trained on some training set. Compared to the white-box MI attacks, the black-box attacks require less~\cite{NasrM2018C} (or even no~\cite{SalemA2018}) auxiliary information about the online MLaaS, which does not release model details but only the queried prediction output. Our focus is on providing privacy against black-box MI attacks.


To prevent membership disclosure, differential privacy (DP)~\cite{DworkC2006C} is a promising technique, which (informally) shields the existence of any arbitrary data sample in a dataset, thereby preserving membership privacy for each member in the training set. There are three broad categories of applying DP into deep learning -- objective perturbation, gradient perturbation and output perturbation -- according to Jayaraman et al.~\cite{JayaramanB2019}. In particular, objective perturbation injects DP noise into the objective function of a learning task~\cite{ZhangJ2012}; gradient perturbation injects DP noise into each epoch during gradient descent~\cite{AbadiM2016}; output perturbation injects DP noise into the elements (edges or nodes) of a trained non-private neural network~\cite{ChaudhuriK2011,WuX2017} or into the prediction results following the sample-and-aggregate mechanism of DP~\cite{PapernotN2018}. Due to the non-convexity of the loss function, applying objective perturbation and output perturbation mainly rely on convexification techniques. For example, Phan et al.~\cite{PhanN2016,PhanN2017} convexify the loss function of convolutional deep belief networks~\cite{LeeH2009}, then inject DP noise into the coefficients via the functional mechanism (originally for non-deep learning tasks)~\cite{ZhangJ2012}. More generally, we could implement output perturbation by training a baseline non-private neural network with a universal convexified loss function~\cite{LoJ2012,DvijothamK2014}, then injecting DP noise into the elements of the trained network.

However, there are some weaknesses that limit the application and performance of existing DP approaches to deep learning. Objective perturbation approaches (combining convexified loss function~\cite{PhanN2016,PhanN2017} and DP objective function~\cite{ZhangJ2012}) only work for a specified learning task -- convolutional deep belief network~\cite{LeeH2009}, which makes it hard to apply it to general deep learning tasks. The gradient perturbation approaches (including using different DP variants and composition theorems) suffer from over-injected noise, mainly since the overall noise injection depends on the number of training epochs, which are usually large in deep learning~\cite{HarderF2020}.
PATE~\cite{PapernotN2018}, the only work implementing sample and aggregate mechanism of DP for output perturbation, works in a special configuration requiring additional publicly available data to assist differentially private aggregation of the distributed learning outputs. The output perturbation framework (universal convexification plus DP noise injection) relies on tight upper bound on the DP noise scale parameter (the global sensitivity). Existing theoretical results~\cite{ChaudhuriK2011,WuX2017} provide loose upper bounds assuming convexity of the loss function. To tighten their results, more conditions should be introduced in addition to convex loss function, such as normalised training sets with binary classes~\cite{ChaudhuriK2011} or smooth loss functions with a decreasing step size during the process of stochastic gradient descent (SGD)~\cite{WuX2017}. These conditions on existing upper bounds are shown in Table~\ref{tab:comparison}.
\begin{table*}[!th]
\centering
\captionsetup{justification=centering}
\caption{Comparison between the Upper Bounds on the $L_{2}$ Global Sensitivity of Trained Model Parameters. \\($\rho$-Lipschitz loss functions, $\lambda$-strongly convex $L_{2}$ regularisers, $\eta$: constant SGD step, $C$: number of classes, \\$\lVert \mathbf{x} \rVert_{2}$: maximum $L_{2}$ norm of a data sample in the training data space, $n$: size of the training set.)}
\label{tab:comparison}
\scalebox{0.9}{
\begin{tabular}{c|c|c|c|c|c|c}
\hline
\multirow{2}{*}{Work} & \multirow{2}{*}{Upper Bound} & \multicolumn{5}{c}{Conditions} \\ \cline{3-7} 
& & Convex loss & Smooth loss & Step size in SGD & Normalised data & Binary classes \\ \hline\hline
Chaudhuri et al.~\cite{ChaudhuriK2011} & $2\rho/\lambda n$ & $\checkmark$ & - & - & $\checkmark$ & $\checkmark$ \\ \hline
Chaudhuri et al.~\cite{ChaudhuriK2011} & $C\lVert \mathbf{x} \rVert_{2}\rho/\lambda n$ & $\checkmark$ & - & - & - & - \\ \hline
Wu et al.~\cite{WuX2017} & $2\rho/\lambda n$ & $\checkmark$ & $\checkmark$ & decreasing & - & - \\ \hline
Wu et al.~\cite{WuX2017} & $2\eta\rho/(1 - (1 - \eta\lambda)^{n})$ & $\checkmark$ & $\checkmark$ & constant & - & - \\ \hline
This Work & $2\rho/\lambda n$ & $\checkmark$ & - & - & - & - \\ \hline
\end{tabular}
}
\end{table*}

\descr{Contributions.} To address the problems in existing DP practice in deep learning, we propose a novel DP framework in the black-box setting. We summarise our main contributions below.
\begin{enumerate}
\setlength\itemsep{0em}
    \item We mathematically analyse a tighter upper bound on the global sensitivity of the trained model parameters, which, like existing upper bounds, assumes a convex loss function, but without other conditions (see Table~\ref{tab:comparison} for a brief comparison). In contrast to existing works, we bound the maximum change of the model parameters by analysing the stability~\cite{ShalevS2014} of a trained model after removing an arbitrary data sample from a training set.
    
    \item We propose a novel framework in the black-box setting, where DP noise is injected into an individual neuron at the output layer of a trained neural network. Specifically, at the training stage (for a baseline non-private model), we apply a universal convexification of the loss function~\cite{LoJ2012,DvijothamK2014} to meet the convexity condition. At the prediction stage (for a differentially private prediction probability vector), we first sample a neuron with the exponential mechanism of DP (Exp-DP)~\cite{McsherryF2007} at the output layer, and then inject DP noise into the sampled neuron, where we scale the noise to the global sensitivity of an individual neuron (bounded by the global sensitivity of the trained model parameters).
    
    \item We empirically compare the privacy-utility trade-off of our framework with existing open-source differentially private stochastic gradient descent (DP-SGD) approaches for classification on six commonly used real-world datasets, following the same experimental configurations as existing studies~\cite{ShokriR2017,JayaramanB2019}. 
    The experimental results show that, when the baseline non-private models have observable privacy leakage under MI attacks, our framework achieves a better privacy-utility trade-off than existing DP-SGD implementations, \reviseA{given an overall privacy budget $\epsilon \leq 1$ for a large number of queries, which is likely to be the size we expect in practice.} 
\end{enumerate}

\section{Related Work}
\label{sec:rw}

Differentially private deep learning can be mainly categorised into objective perturbation, gradient perturbation and output perturbation~\cite{JayaramanB2019}. The difference between the three categories is at the point where we inject DP noise. Objective perturbation injects DP noise into the objective function; gradient perturbation injects DP noise into the process of gradient descent; output perturbation injects DP noise into the output, i.e., the elements of a trained neural network.


There are two works for differentially private deep learning using objective perturbation~\cite{JayaramanB2019}. Phan et al.~\cite{PhanN2016,PhanN2017} replace the non-convex loss function with polynomial approximations via Taylor expansions, then inject DP noise into the coefficients of the approximate function via the functional mechanism~\cite{ZhangJ2012}. However, this framework only works on a specific machine learning task -- convolutional deep belief network~\cite{LeeH2009}, which makes it hard to apply it to general deep learning tasks. Additionally, it is hard to find the optimal polynomial degree to achieve an acceptable trade-off between utility and privacy in practice.

Gradient perturbation is the most used DP technique in deep learning (DP-SGD)~\cite{AbadiM2016}), including an open-source Python library provided by Google's TensorFlow Privacy~\cite{TensorflowPrivacy}. 
DP-SGD perturbs the minibatch stochastic optimisation process, i.e., injecting DP noise into the gradients of each epoch during the training process. Specifically, in each epoch, DP-SGD bounds the global sensitivity of the gradients (the maximum change of the gradients by an individual data samples) by norm clipping. However, since DP-SGD injects noise into all the epochs, \reviseA{the overall injected noise is highly depending on the number of training epochs. If this number is large, which it usually is, DP-SGD would finally inject too much noise.} 

For output perturbation, there are two main streams of techniques. One line of works applies the sample-and-aggregate mechanism of DP to produce differentially private prediction probability vectors. For example, Papernot et al.~\cite{PapernotN2018} propose a framework for differentially private aggregate machine learning, called Private Aggregation of Teacher Ensembles (PATE). Specifically, PATE splits the private dataset into several \reviseA{(at least 100 reported in~\cite{PapernotN2018})} disjoint subsets which are inputs for training sub-models/teacher models. Those teacher models predict the labels of a publicly available data \reviseA{(following the same distribution as the original dataset)}. Then PATE injects DP noise into the count of each predicted label (vote histogram), takes the label having maximum noisy count as the label. Such noisy labelled data are then used to train a student model, which provides differentially private prediction (MLaaS) for new observations. \reviseA{However, PATE may not always be realistic in real-world applications~\cite{HarderF2020}. First, data curators may not have a large amount of private dataset to split, such that each subset is large enough to train a model. Second, data curators may not have any publicly accessible dataset to train the student model. Third, in case there is no such publicly accessible dataset, data curators may use Generative Adversarial Network (GAN)~\cite{GoodfellowI2014} to generate artificial datasets; however, this further introduces additional computational costs, as well as potential impact on utility.}



The other line of work in output perturbation is to inject DP noise into the baseline non-private neural networks trained with a convexified loss function. To do so, we should analyse the upper bound on the global sensitivity of the trained model parameters for noise injection. There are two theoretical results~\cite{ChaudhuriK2011,WuX2017} for this upper bound, subject to the convexity of the loss function. To tighten these bounds, more conditions need to be introduced, such as normalised training data for binary classification tasks~\cite{ChaudhuriK2011} or smooth loss function with decreasing step size during the SGD process~\cite{WuX2017}. However, these additional conditions may not always fit real-world machine learning tasks.

Based on the above analyses, we conclude that existing DP approaches in objective perturbation, gradient perturbation and output perturbation suffer from three problems. First, conditions on objective functions limit the objective perturbation in general deep learning tasks. Second, gradient perturbation does not achieve a satisfactory privacy-utility trade-off due to over-injected noise in each epoch. Third, the utility of output perturbation is not guaranteed because of loose upper bounds on the global sensitivity of the trained model parameters. Therefore, in this paper, we aim to provide a DP framework of output perturbation to control the overall noise injection with a tighter upper bound on global sensitivity of model parameters, achieving a better trade-off between utility and privacy, in general deep learning tasks.

\section{Preliminaries}
\label{sec:preliminaries}
A dataset $X$ is a collection of \emph{rows}: $(\mathbf{x}_{1}, \mathbf{x}_2, \ldots, \mathbf{x}_{n})$, where each $\mathbf{x}_i$ is from some fixed data domain. Two datasets $X$ and $X^{\prime}$ are called neighbouring datasets, denoted $X \sim X^{\prime}$, if one is obtained from the other by adding or removing a single row. Given a dataset $X = (\mathbf{x}_{1}, \dots, \mathbf{x}_{i-1}, \mathbf{x}_{i}, \mathbf{x}_{i+1}, \dots, \mathbf{x}_{n})$, $X^{(-i)}$ represents the dataset $X^{(-i)} = (\mathbf{x}_{1}, \dots, \mathbf{x}_{i-1}, \mathbf{x}_{i+1}, \dots, \mathbf{x}_{n})$, i.e., the neighbouring dataset of $X$ with an arbitrary row $\mathbf{x}_i$ removed. Similarly, $X^{(i)}$ denotes the dataset $X \setminus \{\mathbf{x}_{i}\} \cup \{\mathbf{x}^{\prime}\} = (\mathbf{x}_{1}, \dots, \mathbf{x}_{i-1}, \mathbf{x}^{\prime}, \mathbf{x}_{i+1}, \dots, \mathbf{x}_{n})$, i.e., the row $\mathbf{x}_i$ replaced with an arbitrary row $\mathbf{x}^{\prime}$. Note that the Hamming distance between $X$ and $X^{(-i)}$, $d(X, X^{(-i)}) = 1$ and datasets $X$ and $X^{(i)}$ are \emph{not} neighbouring datasets according to our definition. 

\subsection{Differential Privacy} 
Informally, differential privacy (DP)~\cite{DworkC2006C} ensures that two outputs, produced by a randomised algorithm on two neighbouring datasets, are almost indistinguishable. 
\begin{definition}[$\epsilon$-DP~\cite{DworkC2006C}]
\label{def:dp}
A randomised mechanism $\mathcal{T}$ is $\epsilon$-differentially private if for all neighbouring datasets $X$ and $X^{(-i)}$, $i \sim U(n)$, and for all outputs $S \in \text{Range}(\mathcal{T})$, $\mathcal{T}$ satisfies: 
\begin{equation}
\label{eq:dp}
\Pr[\mathcal{T}(X) = S]\leq \exp(\epsilon) \times \Pr[\mathcal{T}(X^{(-i)}) = S],
\end{equation}
where $\epsilon > 0$ is the privacy budget.
\end{definition}

Two prototypical $\epsilon$-DP mechanisms considered in this paper are the Laplace mechanism (Lap-DP)~\cite{DworkC2006C} and the exponential mechanism (Exp-DP)~\cite{McsherryF2007}. The former adds additive noise drawn from the Laplace distribution of scale $\Delta f/\epsilon$ to the numeric computation $f(\cdot)$, where $\Delta f$ is the $L_{1}$ global sensitivity of $f$. The $L_{p}$ global sensitivity of $f$ is:
\begin{equation}
\label{eq:gs}
\Delta_{p} f = \max_{\forall X \in \mathcal{D},d(X,X^{(-i)})=1}\lVert f(X) - f(X^{(-i)}) \rVert_{p}.
\end{equation}
where $\lVert \cdot \rVert_{p}$ is the $L_{p}$-norm. For simplicity, in this paper, $\Delta f$ denotes the $L_{1}$ global sensitivity.

The Exp-DP mechanism is a weighted sampling scheme to generate the output from a set of (arbitrary) candidates, and is well-suited for non-numeric computations. The mechanism simply outputs a given candidate $r$ from a set of candidates $R$ computed over a dataset $X$, with probability proportional to $\exp(\epsilon q(X,r)/2\Delta q)$, where $q(X,r)$ is the score/quality function of the candidate $r$, and $\Delta q$ is the maximum possible change in $q$ over all neighbouring datasets.


An important property of differential privacy is that the mechanisms compose~\cite{DworkC2014}: 
\begin{theorem}[Sequential Composition~\cite{DworkC2014}]
\label{thm:Lap-DP_composition}
Let $M_{i}(X)$ provide $\epsilon_{i}$-DP. Then the sequence of $M_{i}(X)$ provides $\sum_{i}\epsilon_{i}$-DP.
\end{theorem}


The recently proposed $\epsilon$-Gaussian DP ($\epsilon$-GDP)~\cite{DongJ2019}, injecting noise following a Gaussian distribution $\mathcal{N}(0, (\sfrac{\Delta f}{\epsilon})^{2})$, is a relaxation of traditional $\epsilon$-DP. We have
\begin{theorem}~\cite{DongJ2019}
\label{thm:gdp}
A mechanism is $\epsilon$-GDP if and only if it is $(\epsilon, \delta(\epsilon))$-DP, $\forall \epsilon > 0$, where $\delta(\epsilon) = \Phi(-1 + \sfrac{\epsilon}{2}) - \exp(\epsilon)\Phi(-1 - \sfrac{\epsilon}{2})$, $\Phi(\cdot)$ is the cumulative distribution function of $\mathcal{N}(0, 1)$.
\end{theorem}

The composition of GDP is as follows.
\begin{theorem}[Composition of GDP~\cite{DongJ2019}]
\label{thm:GDP_composition}
The $n$-fold composition of $\epsilon_{i}$-GDP is $\sqrt{\sum_{i=1}^{n}\epsilon_{i}^{2}}$-GDP, where each $\epsilon_{i}$-GDP injects Gaussian noise following $\mathcal{N}(0, (\sfrac{\Delta f}{\epsilon_{i}})^{2})$.
\end{theorem}

\subsection{Machine Learning Basics}
Let $A$ be a learning algorithm, $X = (\mathbf{x}_{1}, \dots, \mathbf{x}_{n})$ be a training set drawn from a distribution $\mathcal{D}$ ($X \sim \mathcal{D}^{n}$), $W_{X} (= (\omega_{1}, \dots, \omega_{|E|}) \in \mathbb{R}^{|E|}) = A(X)$ be the output of the learning algorithm $A$ on the training set $X$, i.e., trained model parameters, where $|E|$ is the number of model parameters. Once the model parameters are set after training, we predict the class label $y$ of an observation $\mathbf{x}$ as $y = \argmax \{p_{1}, \dots, p_{C}\}$, where $\mathbf{p} = \{p_{1}, \dots, p_{C}\} = h(W_{X}, \mathbf{x})$, $h$ is the learnt hypothesis and $C$ is the number of classes.

We use a loss function $l(W_{X},\mathbf{x}_{i})$ on an observation $\mathbf{x}_{i} \in X \sim \mathcal{D}^{n}$ to measure the empirical error of $W_{X}$ on $\mathbf{x}_{i}$. Note that each $\mathbf{x}_{i}$ is a composition of a set of features and an accompanying class label. Accordingly, we have the following equation to measure the training error of the learning algorithm $A$ on the dataset $X$. The learning algorithm $A$ aims to minimise the training error. 
\begin{equation}
\label{eq:loss}
    L_{X}(W_{X}) = \frac{1}{|X|}\sum_{i=1}^{|X|}l(W_{X}, \mathbf{x}_{i})
\end{equation}

\descr{Measurement for overfitting.} A fundamental issue to avoid in machine learning is overfitting of a learning algorithm. 
One way to measure overfitting of a learning algorithm is the notion of On-Average-Replace-One stability of a model~\cite{ShalevS2014}. That is, replacing a single data sample does not result in a large difference on the value of the loss function.
\begin{definition}[On-Average-Replace-One Stability~\cite{ShalevS2014}]
\label{def:replace_stable}
Let $\sigma:\mathbb{N} \to \mathbb{R}$ be a monotonically decreasing function. We say that a learning algorithm $A$ is on-average-replace-one stable with rate $\sigma(n)$ if for every distribution $\mathcal{D}$,
\begin{equation}
\label{eq:replace_stable}
    \mathbb{E}_{(X, \mathbf{x}^{\prime}) \sim \mathcal{D}^{n+1}, i \sim U(n)}[l(W_{X^{(i)}}, \mathbf{x}_{i}) - l(W_{X}, \mathbf{x}_{i})] \leq \sigma(n),
\end{equation}
where $X^{(i)} = X \setminus \{\mathbf{x}_{i}\} \cup \{\mathbf{x}^{\prime}\}$, 
and $U(n)$ is the uniform distribution over $[n]$.
\end{definition}

We extend On-Average-Replace-One stability to On-Average-Remove-One stability 
to fit the neighbouring dataset requirement of Definition~\ref{def:dp}. 
\begin{definition}[On-Average-Remove-One (OARO) Stability]
\label{def:remove_stable}
Given a monotonically decreasing function $\sigma:\mathbb{N} \to \mathbb{R}$, an arbitrary distribution $\mathcal{D}$, then a learning algorithm $A$ is on-average-remove-one stable with rate $\sigma(n)$ if
\begin{equation}
\label{eq:remove_stable}
    \mathbb{E}_{X \sim \mathcal{D}^{n}, i \sim U(n)}[l(W_{X^{(-i)}}, \mathbf{x}_{i}) - l(W_{X}, \mathbf{x}_{i})] \leq \sigma(n),
\end{equation}
where $U(n)$ is the uniform distribution over $[n]$.
\end{definition}
Based on Inequality~\eqref{eq:remove_stable}, a less $\sigma(n)$ indicates higher OARO stability.


\descr{Topology of deep neural networks.} A deep neural network is a graph, $G = (V, E)$, formed by connecting $T+1$ layers. A layer $V_{t}$ $(t \in [0, T])$ is a set of disconnected vertices, $V_{t} = \{v_{t,1}, v_{t,2}, \dots, v_{t, |V_{t}|}\}$. Each vertex is a neuron in the neural network, one of the neurons serves as the bias term. The vertices set $V$ is a union of disjoint layers, $V = \bigcup_{t=0}^{T} V_{t} $ and the edges set $E$ is the set of weighted edges connecting vertices in two adjacent layers. For example, an edge $e_{t,i,j}$ (with a weight $\omega_{t,i,j}$) connects $v_{t-1,j}$ and $v_{t,i}$. We call $V_{0}$ input layer, $V_{t}$ $(t \in [1, T-1])$ hidden layers and $V_{T}$ output layer. Note that $|V_{0}| = m + 1$, where $m$ is the dimension of a data sample, $|V_{T}| = C$, where $C$ is the number of classes, $|E| = |W_{X}|$ is the number of model parameters.

When using a trained neural network (trained model parameters/weights and given topology) to make prediction for a given data sample $\mathbf{x} = (x_{1}, \dots, x_{n})$, we calculate the output of each vertex/neuron, $a_{t,i}$ at layer $V_{t}$ as follows.
\begin{equation}
\label{eq:neuron}
\begin{cases}
    z_{t,i} & = \sum_{\forall e_{t,i,j}} \omega_{t,i,j}a_{t-1,j}, \\
    a_{t,i} & = \phi(z_{t,i}),
\end{cases}
\end{equation}
where $\phi(\cdot)$ is an activation function for the neurons, $a_{0,i} = x_{i}$ and $a_{t-1,|V_{t-1}|}$ is also known as the bias term. At the output layer, the final prediction probability vector of the given data sample $\mathbf{x}$ would be $\mathbf{p} = (a_{T,1}, \dots, a_{T,C})$, where $\sum_{i=1}^{C} a_{T,i} = 1$. In our experiments in Section~\ref{sec:exp}, $\phi(\cdot)$ is Tanh for $t < T$, $\phi(\cdot)$ is Softmax for $t = T$. 


Furthermore, we also use the concepts of \textit{convex set}, \textit{convex function}, $\lambda$-\textit{strong convexity} and $\rho$-\textit{Lipschitzness} in this paper. We follow the standard definitions of these concepts.


\begin{definition}[Convex set~\cite{ShalevS2014}]
$X$ is a convex set if and only if for any two vector $\mathbf{u}$, $\mathbf{v} \in X$ and an $\alpha \in [0, 1]$, we have $\alpha\mathbf{u} + (1 - \alpha)\mathbf{v} \in X$.
\end{definition}

\begin{definition}[Convex function~\cite{ShalevS2014}]
\label{def:convex}
A function $f: X \to \mathbb{R}$ is convex if $X$ is convex and for any two vectors $\mathbf{u}$ and $\mathbf{v}$ in $X$, we have 
\begin{equation}
    f(\alpha\mathbf{u} + (1 - \alpha)\mathbf{v}) \leq \alpha f(\mathbf{u}) + (1 - \alpha)f(\mathbf{v}).
\end{equation}
\end{definition}


\begin{definition}[$\lambda$-Strong convexity~\cite{ShalevS2014}]
\label{def:strong_convex}
A convex function $f: X \to \mathbb{R}$ is $\lambda$-strongly convex if for any two vectors $\mathbf{u}$ and $\mathbf{v}$ in $X$ and $\alpha \in [0, 1]$, we have
\begin{equation}
    f(\alpha\mathbf{u} + (1 - \alpha)\mathbf{v}) \leq \alpha f(\mathbf{u}) + (1 - \alpha)f(\mathbf{v}) - \frac{\lambda}{2}\alpha(1 - \alpha)\lVert \mathbf{u} - \mathbf{v} \rVert^{2}_{2}.
\end{equation}
\end{definition}

\begin{definition}[$\rho$-Lipschitzness~\cite{ShalevS2014}]
\label{def:lip}
A function $f: X \to \mathbb{R}^{d}$ is $\rho$-Lipschitz over $X$ if for any two vectors $\mathbf{u}$ and $\mathbf{v}$ in X, we have
\begin{equation}
    \lVert f(\mathbf{u}) - f(\mathbf{v}) \rVert \leq \rho\lVert \mathbf{u} - \mathbf{v} \rVert.
\end{equation}
\end{definition}


\subsection{Membership Inference Attacks}
\label{subsec:mia}
The goal of Membership Inference (MI) Attacks is to infer the membership of a given data sample by the prediction output of the data sample. In this paper we specify MI attacks as black-box MI attacks implemented via the shadow model technique by Shokri et al.~\cite{ShokriR2017}. That is, the MI adversaries only have access to the distribution of the target/private dataset and the prediction output of a given data sample. 

For simplicity, a black-box MI attack via shadow models is a function $\mathcal{A}: \mathcal{D}^{kn+1} \to \{0, 1\}$. The adversary performs a shadow models-based MI attack in three steps. Firstly, the adversary trains $k$ shadow models with shadow datasets $X^{\prime}_{i} \sim \mathcal{D}^{n}$, $i \in [1, k]$, mimicking the behaviour of the target model trained with the target/private dataset $X$. Then the adversary uses the $k$ shadow models to make predictions for training data (members) and test data (non-members). These prediction vectors will then be the training data to train an attack model (a binary classifier). Finally, for a given data sample $\mathbf{x} \sim \mathcal{D}$, the adversary queries the prediction probability vector $\mathbf{p}$ of $\mathbf{x}$ from the target model, then feeds $\mathbf{p}$ to the trained attack model to predict the membership ($\{0,1\}$) of $\mathbf{x}$.

Table~\ref{tab:notations} summarises the notations used in this paper.
\begin{table}[!ht]
\centering
\caption{Summary of Notations.}
\scalebox{0.9}{
\begin{tabular}{c|l}
\hline
Notation & Description \\
\hline\hline
 $A$ & Learning algorithm \\
\hline
 $C$ & Number of classes \\
\hline
 $\mathcal{D}$ & Data distribution \\
\hline
 $\epsilon$ & Privacy budget \\
\hline
 $e_{t,i,j}$ & Edge connecting $v_{t-1,j}$ and $v_{t,i}$ \\
\hline
 $G = (V, E)$ & Neural network \\
\hline
 $\lambda$ & Strongly convex constant \\
\hline
 $l(W_{X}, \mathbf{x})$ & Loss function on $\mathbf{x}$ \\
\hline
 $L_{X}(W_{X})$ & Training error on $X$ \\
\hline
 $m$ & Dimension of the training set \\
\hline
 $n$, $|X|$ & Size of the training set \\
\hline
 $\rho$ & Lipschitzness constant \\
\hline
 $U(n)$ & Uniform distribution over $[1, n]$ \\
\hline
 $V_{t}$ & Neurons set at layer $t$ \\
\hline
 $W_{X}$, $A(X)$ & Model parameters trained on $X$ \\
\hline
 $\Delta_{2} W$ & $L_{2}$ global sensitivity of weights vector \\
\hline
 $\Delta \omega$ & $L_{1}$ global sensitivity of an individual weight \\
\hline
 $\omega_{t,i,j}$ & Weight on edge $e_{t,i,j}$ \\
\hline
 $v_{t,i}$ & $i$th neuron at $t$th layer in $G$ \\
\hline
 $X$ & Training set \\
\hline
 $X^{(-i)}$ & Neighbouring dataset of $X$ \\
\hline
 $\mathbf{x}$ & Data sample (vector) of $X$ \\
\hline
 $\mathbf{x} \sim \mathcal{D}$ & $\mathbf{x}$ drawn from distribution $\mathcal{D}$ \\
\hline
 $\Delta z_{T}$ & $L_{1}$ global sensitivity of a neuron (output layer) \\
\hline
 $\lVert \mathbf{a}, \mathbf{b} \rVert_{p}$ & $L_{p}$ distance between $\mathbf{a}$ and $\mathbf{b}$ \\
\hline
 $|S|$ & Cardinality of a set $S$ \\
\hline
\end{tabular}
}
\label{tab:notations}
\end{table}

\section{Differential Privacy via Output Perturbation and Theoretical Results}
\label{sec:main}
In this section, we show our differentially private algorithm, then mathematically analyse the properties of the proposed approach including global sensitivity of model parameters and the effect of differential privacy against membership inference attacks.

\subsection{Idea Overview}
\label{sec:overview}
\reviseA{Based on our discussion in the related works section (Section~\ref{sec:rw}), an output perturbation-based DP solution has one potential benefit over gradient and input perturbation-based approaches. In the latter two approaches, the entire privacy budget $\epsilon$ is consumed for model training. Any subsequent use of the trained model does not further consume the privacy budget due to the post-processing property of differential privacy. This, however, means that the utility offered is fixed for any number of queries; whether we ask 10 queries or 10,000 queries (a query means one data sample given as input to the model; the model in turn returns the corresponding probability vector). On the other hand, the output perturbation approach depletes budget per query to the trained model. This means that if the trained model is used for a fixed number of queries (as opposed to an unlimited number of queries), then we can offer better utility by virtue of allocating larger chunks of the budget per query. This motivates our focus on output perturbation.} 

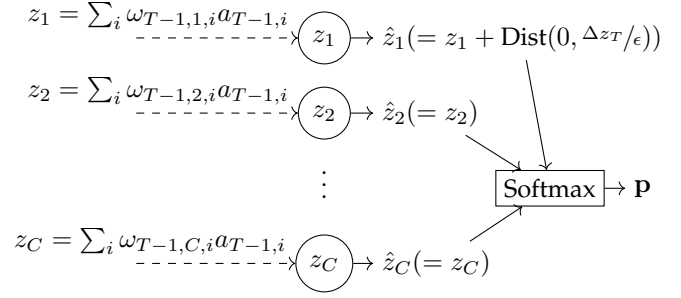
\begin{figure}[!ht]
    \centering
    \begin{tikzpicture}[xscale=0.5, yscale=0.5]
    \node [draw, circle, name=a1] at (0,0) {$z_{1}$};
    \node [right, name=z1] at (1.3,0) {$\hat{z}_{1} (= z_{1} + \text{Dist}(0,\sfrac{\Delta z_{T}}{\epsilon}))$};
    \path[->,draw] (a1) to (z1);
    \path[->, dashed] (-5,0) edge node[above left, pos=1] {$z_{1} = \sum_{i}\omega_{T-1,1,i}a_{T-1,i}$} (a1);
    
    \node [draw, circle, name=a2] at (0,-2) {$z_{2}$};
    \node [right, name=z2] at (1.3,-2) {$\hat{z}_{2} (= z_{2})$};
    \path[->,draw] (a2) to (z2);
    \path[->, dashed] (-5,-2) edge node[above left, pos=1] {$z_{2} = \sum_{i}\omega_{T-1,2,i}a_{T-1,i}$} (a2);
    
    \node [draw, circle, name=an] at (0,-6) {$z_{C}$};
    \node [right, name=zn] at (1.3,-6) {$\hat{z}_{C} (= z_{C})$};
    \path[->,draw] (an) to (zn);
    \path[->, dashed] (-5,-6) edge node[above left, pos=1] {$z_{C} = \sum_{i}\omega_{T-1,C,i}a_{T-1,i}$} (an);
    
    \path (a2) -- (an) node [midway, sloped] {$\dots$};
    
    \node [draw, rectangle, name=c] at (6,-4) {$\text{Softmax}$};
    \path[->, draw] (z1) to (c);
    \path[->, draw] (z2) to (c);
    \path[->, draw] (zn) to (c);
    
    \node [name=p, right] at (8,-4) {$\mathbf{p}$};
    \path[->, draw] (c) to (p);
    \end{tikzpicture}
    \caption{Idea Overview (assume $z_{1}$ is the sampled neuron).}
    \label{fig:overview}
\end{figure}

In this section, we propose an output perturbation-based DP algorithm for general deep learning tasks. To provide a better privacy-utility trade-off, we inject DP noise into a randomly sampled neuron at the output layer. 
Figure~\ref{fig:overview} depicts our general idea, where neuron $z_{1}$ is the randomly sampled neuron and the DP noise follows the distribution $\text{Dist}(0, \sfrac{\Delta z_{T}}{\epsilon})$. \reviseA{
We shall use both the Laplace and the Gaussian mechanisms as examples of this distribution in our experiments, providing $\epsilon$-DP and $\epsilon$-GDP (or $(\epsilon, \delta(\epsilon))$-DP) guarantees, respectively.}

\begin{algorithm}[!t]
\small
\caption{DP Prediction Probability Vector for Non-Private Deep Learning Models.}
\label{alg:dpdnn}
\SetKwInOut{Input}{Input}
\SetKwInOut{Output}{Output}
\Input{
    $G = (V, E)$: an artificial neural network; \\
    $C$: number of classes; \\
    $W_{X} = \{\omega_{t,i,j} | t \in [1, T], i \in [1, |V_{t}|], j \in [1, |V_{t-1}|]\}$: trained model parameters on dataset $X$; \\
    $\phi(\cdot)$: activation function at hidden layers; \\
    $\epsilon_{\text{sampling}}$: privacy budget for sampling a neuron; \\
    $\epsilon_{\text{neuron}}$: privacy budget for injecting noise into a neuron; \\
    $\Delta z_{T}$: $L_{1}$ global sensitivity of the neurons at layer $T$; \\
    $\Delta p$: $L_{1}$ global sensitivity of the prediction probability vector; \\
    $\mathbf{x}$: a data sample (query).
}
\Output{
  $\mathbf{p} = (p_{1}, \dots, p_{C})$: differentially private prediction probability vector.
}
\BlankLine
$\mathbf{a}_{0}$ $\gets$ $\mathbf{x}$\;
\For{$t$ $\gets$ $0$ to $T-1$}{
    \For{$i$ $\gets$ $1$ to $|V_{t+1}|$}{
        \uIf{$t \neq T-1$}{
            $a_{t+1,i}$ $\gets$ $\phi(\sum_{j=1}^{|V_{t}|}\omega_{t+1,i,j}a_{t,j})$\;
        }
        \Else{
            $z_{t+1,i}$ $\gets$ $\sum_{j=1}^{|V_{t}|}\omega_{t+1,i,j}a_{t,j}$\;
        }
    }
}
$\mathbf{p}$ $\gets$ $\text{Softmax}(z_{T,1}, \dots, z_{T,C})$\;
$v$ $\gets$ $\text{Exp-DP}(\{z_{T,1}, \dots, z_{T,C}\}, \Pr[v] \propto \exp(\sfrac{\epsilon_{\text{sampling}} p_{v}}{2\Delta p}))$\;
$\hat{z}_{T,v}$ $\gets$ $z_{T,v} + \text{Dist}(0, \sfrac{\Delta z_{T}}{\epsilon_{\text{neuron}}})$\;
$(\hat{z}_{T,1}, \dots, \hat{z}_{T,C})$ $\gets$ $(z_{T,1}, \dots, \hat{z}_{T,v}, \dots, z_{T,C})$\;
$\mathbf{p}$ $\gets$ $\text{Softmax}(\hat{z}_{T,1}, \dots, \hat{z}_{T,C})$\;
\Return: $\mathbf{p}$\;
\end{algorithm}

Specifically, our DP algorithm works in three steps. Firstly, for a given data sample, we feed this data sample to a trained non-private neural network to calculate the values of the neurons at the output layer. Next, we inject DP noise into a randomly sampled neuron at the output layer. After noise injection, we apply Softmax function on the noisy neuron vector to produce a differentially private probability vector. Algorithm~\ref{alg:dpdnn} shows the implementation details, where Line 1 to Line 7 are the first step, Line 8 to Line 10 are the second step, and Line 11 to Line 13 are the third step. \reviseA{Note that, Algorithm~\ref{alg:dpdnn} only introduces marginally more time than the non-private neural network: Lines 9 and 10 to the non-private neural networks. Sampling (Line 9) is efficiently implementable even with a large number of classes. The injected noise does not depend on the number of classes and the number of training data samples.}

\reviseA{Also, since noise is not injected into the rest of the neurons in the neural network, the output perturbation technique is useful only in the black-box setting, where model parameters are not released.} The key point of implementing Algorithm~\ref{alg:dpdnn} is to find a tight upper bounds on $\Delta z_{T}$ for deep learning models. In the following sections, we show our upper bound on $\Delta z_{T}$ by analysing the upper bound on the global sensitivity of the model parameters, $\Delta_{2} W$ (which is tighter than existing results~\cite{ChaudhuriK2011,WuX2017} subject to the convexity of the loss function). We then show the DP guarantee of Algorithm~\ref{alg:dpdnn} and the effect of DP against the black-box MI attacks.


\subsection{Analysis of the Upper Bound on Global Sensitivity}
\label{sec:global_sens}
This section analyses the upper bounds on the $L_{2}$ global sensitivity of trained model parameters and the $L_{1}$ global sensitivity of an individual neuron at the output layer. We analyse these upper bounds based on the fundamental properties of convex and strongly convex functions.

\begin{lemma}~\cite{ShalevS2014}
\label{lm:l2_norm_regulariser}
The $L_{2}$-norm regulariser $2\lambda\lVert W_{X} \rVert^{2}_{2}$ is $\lambda$-strongly convex.
\end{lemma}

\begin{lemma}~\cite{ShalevS2014}
\label{lm:lambda_strong}
Let function $h_{i}$ be convex, function $g$ be $\lambda$-strongly convex. Their linear composition $f = \frac{1}{n}\sum_{i=1}^{n}h_{i} + g$ is $\lambda$-strongly convex.
\end{lemma}

\begin{lemma}~\cite{ShalevS2014}
\label{lm:minimiser}
Let $\mathbf{u}$ be a minimiser of a $\lambda$-strongly convex function $f$ ($f^{\prime}(\mathbf{u}) = 0$), then for any $\mathbf{v}$, we have
\begin{equation}
    f(\mathbf{v}) - f(\mathbf{u}) \geq \frac{\lambda}{2}\lVert \mathbf{v}-\mathbf{u} \rVert^{2}_{2}.
\end{equation}
\end{lemma}




Based on Lemma~\ref{lm:l2_norm_regulariser}, Lemma~\ref{lm:lambda_strong} (when $h$ is the loss function and $g$ is the $L_{2}$-norm regulariser) and Lemma~\ref{lm:minimiser}, we have the following theorem to find the upper bound on the On-Average-Remove-One (OARO) stability (Definition~\ref{def:replace_stable}) of the loss function and the global sensitivity of the model parameters, where we measure the global sensitivity of the model parameters by $L_{2}$-norm (a.k.a. $L_{2}$-sensitivity~\cite{DworkC2014}). 

\begin{theorem}
\label{thm:global_sensitivity}
Given a convex and $\rho$-Lipschitz loss function and a $\lambda$-strongly convex regulariser, the upper bound on the $L_{2}$ global sensitivity of the model parameters (of the neural network) $\Delta_{2} W$ is $\sfrac{2\rho}{\lambda n}$; the On-Average-Remove-One stability of the loss function $l(W_{X}, \mathbf{x})$ is bounded by $\sfrac{2\rho^{2}}{\lambda n}$.
\end{theorem}
\begin{proof}

Based on Lemma~\ref{lm:l2_norm_regulariser}, Lemma~\ref{lm:lambda_strong} and Lemma~\ref{lm:minimiser}, we have
\begin{align}
    & \lVert W_{X^{(-i)}} - W_{X} \rVert^{2}_{2} \nonumber \\
    \leq & \frac{2}{\lambda}\left(L_{X}(W_{X^{(-i)}}) + \frac{\lambda}{2}\lVert W_{X^{(-i)}} \rVert^{2}_{2}\right) \nonumber \\
    & - \frac{2}{\lambda}\left(L_{X}(W_{X}) + \frac{\lambda}{2}\lVert W_{X} \rVert^{2}_{2}\right) \label{subeq:lambda_strong} \\
    = & \frac{2}{\lambda}\left(L_{X^{(-i)}}(W_{X^{(-i)}}) + \frac{\lambda}{2}\lVert W_{X^{(-i)}} \rVert^{2}_{2}\right) \nonumber \\
    & - \frac{2}{\lambda}\left(L_{X^{(-i)}}(W_{X}) + \frac{\lambda}{2}\lVert W_{X} \rVert^{2}_{2}\right) \nonumber \\
    & + \frac{2}{\lambda n}\left(l(W_{X^{(-i)}}, \mathbf{x}_{i}) - l(W_{X}, \mathbf{x}_{i})\right) \label{subeq:change_measure} \\
    \leq & \frac{2}{\lambda n}\left(l(W_{X^{(-i)}}, \mathbf{x}_{i}) - l(W_{X}, \mathbf{x}_{i})\right) \label{subeq:minimiser} \\
    \leq & \frac{2\rho}{\lambda n} \lVert W_{X^{(-i)}} - W_{X} \rVert_{2}, \label{subeq:rho_lip}
\end{align}
where Inequality~\eqref{subeq:lambda_strong} is based on Lemma~\ref{lm:minimiser}; we change the measurement domain for the training error in Equation~\eqref{subeq:change_measure}, such that one more term is added for each training error; since $W_{X^{(-i)}}$ is the minimiser of $L_{X^{(-i)}}(W_{X^{(-i)}}) + \frac{\lambda}{2}\lVert W_{X^{(-i)}} \rVert^{2}_{2}$, we have Inequality~\eqref{subeq:minimiser}; Inequality~\eqref{subeq:rho_lip}, $\forall \mathbf{x}_{i} \in X$ is a result of $\rho$-Lipschitz loss function. Because $\lVert W_{X^{(-i)}} - W_{X} \rVert_{2} \geq 0$, $i \sim U(|X|)$, we immediately have 
\begin{equation}
\label{eq:global_sensitivity_weights}
    \Delta_{2} W = \lVert W_{X^{(-i)}} - W_{X} \rVert_{2} \leq \frac{2\rho}{\lambda n}.
\end{equation}

Furthermore, since the loss function $l(W_{X}, \mathbf{x})$ is $\rho$-Lipschitz, we have
\begin{align}
    l(W_{X^{(-i)}}, \mathbf{x}_{i}) - l(W_{X}, \mathbf{x}_{i}) \leq \rho \lVert W_{X^{(-i)}} - W_{X} \rVert_{2} \leq \frac{2\rho^{2}}{\lambda n}.
\end{align}
Since this inequality holds for any $X$ and $\mathbf{x}_{i}$ ($i \sim U(|X|)$), we then have
\begin{align}
\label{eq:upper_loss}
    \mathbb{E}_{(X, \mathbf{x}_{i}) \sim \mathcal{D}^{n+1}, i \sim U(n)}[l(W_{X^{(-i)}}, \mathbf{x}_{i}) - l(W_{X}, \mathbf{x}_{i})] \leq \frac{2\rho^{2}}{\lambda n}.
\end{align}

Combining Inequality~\eqref{eq:global_sensitivity_weights} and Inequality~\eqref{eq:upper_loss}, concludes the proof.
\end{proof}

We note that Chaudhuri et al.~\cite{ChaudhuriK2011} provide a similar result to Theorem~\ref{thm:global_sensitivity}. However, their upper bound on the $L_{2}$ global sensitivity of the model parameters is $\sfrac{2\rho}{\lambda n}$ under the condition of a binary classification task and a normalised training set ($\lVert \mathbf{x}_{i} \rVert \leq 1$). In addition, Wu et al.~\cite{WuX2017} also provide the same upper bound on the global sensitivity of the model parameters as ours. However, their result needs the loss function to be $\beta$-smooth with a decreasing step size during the process of stochastic gradient descent. Once we remove these additional conditions, the two upper bounds become loose. Table~\ref{tab:comparison} in Section~\ref{sec:introduction} shows a brief comparison between these upper bonds and their conditions. Compared to \cite{ChaudhuriK2011} and \cite{WuX2017}, when achieving the same tight bound, our result only relies on the convexity of the loss function but not the additional conditions introduced in \cite{ChaudhuriK2011,WuX2017}.

Since $\Delta_{2} W$ is the $L_{2}$ global sensitivity, we have $(\Delta_{2} W)^{2} = \sum_{i=1}^{|W_{X}|}(\Delta \omega_{i})^{2}$, where $|W_{X}|$ is the number of edges in a deep neural network. Since each $\Delta \omega_{i}$ has the same upper bound $\Delta \omega$, we have the global sensitivity ($L_{1}$-norm) of an individual model parameter
\begin{equation}
\label{eq:global_sensitivity_weight}
\Delta \omega = \frac{\Delta_{2} W}{\sqrt{|W_{X}|}} \leq \frac{2 \rho}{\lambda n \sqrt{|W_{X}|}}.
\end{equation}


Once we have the $L_{1}$ global sensitivity of each model parameter/weight, we can further analyse the upper bound on the global sensitivity of the neuron (input to the Softmax function) at the output layer.

\begin{corollary}
\label{cr:global_sensitivity_output}
Given the $L_{1}$ global sensitivity of an individual model parameter $\Delta \omega$, a fully connected $(T+1)$-layer neural network, $G = (V, E)$, where the output layer $V_{T}$ has no activation function, the activation function at the hidden layers is bounded by $a_{u}$, the $L_{1}$ global sensitivity of a neuron $v_{T}$ at the output layer is bounded by $a_{u}|V_{T-1}|\Delta \omega$.
\end{corollary}
\begin{proof}
Let $z_{T}$ be the value of an arbitrary neuron at the output layer $v_{T}$, then we have $z_{T} = \sum_{j=1}^{|V_{T-1}|}a_{T-1,j}\omega_{T,j}$, where $|V_{T-1}|$ is the incoming degree of $v_{T}$ in a fully connected neural network. Since the activation function is bounded by $a_{u}$. We achieve the maximum difference between $z_{T}$ and $z_{T}^{(-i)}$ when $\omega_{T,j} > 0 > \omega_{T,j}^{(-i)}$ and $a_{T-1,j} = a_{T-1,j}^{(-i)} = a_{u}$, $\forall j \sim U(|V_{T-1}|)$, that is
\begin{align}
\label{eq:global_neuron}
    \Delta z_{T} & = \sum_{j=1}^{|V_{T-1}|}a_{u}\omega_{T,j} - \sum_{j=1}^{|V_{T-1}|}a_{u}\omega_{T,j}^{(-i)} \nonumber \\
    & \leq a_{u}\sum_{j=1}^{|V_{T-1}|}\left(\omega_{T,j} - \omega_{T,j}^{(-i)}\right) \nonumber \\
    & \leq a_{u}|V_{T-1}|\Delta \omega
\end{align}
as required.
\end{proof}

In practice, some commonly used activation functions, such as, Tanh, Sigmoid, Binary step and Gaussian functions, provide $a_{u} = 1$. In our experiments (Section~\ref{sec:exp}), we use Tanh function as the activation function in the hidden layers, following existing works~\cite{ShokriR2017,SalemA2018,YeomS2018,JayaramanB2019} in MI attacks.


\begin{corollary}
\label{cr:sensitivity_softmax}
Given the global sensitivity of an individual neuron at the output layer of a neural network $\Delta z_{T}$, the upper bound on the global sensitivity of $p \in \mathbf{p}$, $\Delta p$, is $\min\{\exp(2\Delta z_{T}) - 1, 1\}$, where $\mathbf{p}$ is the prediction probability vector provided by the Softmax function.
\end{corollary}
\begin{proof}
For an arbitrary neuron $v$ at the output layer, we have $p_{v} = \frac{\exp(z_{T,v})}{\sum_{j=1}^{C}\exp(z_{T,j})}$, where $C$ is the number of classes. Since $z_{T,j} - \Delta z_{T} \leq z_{T,j}^{(-i)} \leq z_{T,j} + \Delta z_{T}$, the global sensitivity of $p$ is
\begin{align}
    \Delta p = & \sup\left|p_{v}^{(-i)} - p_{v}\right| \nonumber \\
    = & \sup\left|\frac{\exp(z_{T,v}^{(-i)})}{\sum_{j=1}^{C}\exp(z_{T,j}^{(-i)})} - \frac{\exp(z_{T,v})}{\sum_{j=1}^{C}\exp(z_{T,j})}\right| \nonumber \\
    = & \frac{\exp(z_{T,v} + \Delta z_{T})}{\exp(z_{T,v} + \Delta z_{T}) + \sum_{j \neq v}\exp(z_{T,j} - \Delta z_{T})} \nonumber \\
    & \quad - \frac{\exp(z_{T,v})}{\sum_{j=1}^{C}\exp(z_{T,j})} \nonumber \\
    = & \frac{\exp(z_{T,v})}{\exp(z_{T,v}) + \sum_{j \neq v}\exp(z_{T,j} - 2\Delta z_{T})} \nonumber \\
    & \quad - \frac{\exp(z_{T,v})}{\sum_{j=1}^{C}\exp(z_{T,j})} \nonumber \\
    \leq & \frac{\exp(z_{T,v})}{\sum_{j=1}^{C}\exp(z_{T,j} - 2\Delta z_{T})} - \frac{\exp(z_{T,v})}{\sum_{j=1}^{C}\exp(z_{T,j})} \nonumber \\
    = & (\exp(2 \Delta z_{T}) - 1) \times \frac{\exp(z_{T,v})}{\sum_{j=1}^{C}\exp(z_{T,j})} \nonumber \\
    < & \exp(2 \Delta z_{T}) - 1.
\end{align}
Since both $p_{v}$ and $p_{v}^{(-i)}$ are less than $1$, we have $\Delta p = \min\{\exp(2 \Delta z_{T}) - 1, 1\}$ to conclude this corollary.
\end{proof}

\reviseA{Note that, when $\Delta z_{T} > 0.5 \ln{2}$, $\Delta p$ becomes trivial, i.e., $\Delta p = 1$, which is the maximum change for a probability $p_{v}$ of neuron $v$ at the output layer). However, based on the way the Exp-DP behaves (in Algorithm~\ref{alg:dpdnn}), even if the bound on $\Delta p$ is trivial, the highest probability neuron will still have a relatively high probability of being sampled. That is, Algorithm~\ref{alg:dpdnn} still guarantees the quality of the sampled neuron. Moreover, in Table~\ref{tab:sensitivity} in Section~\ref{sec:exp_dp}, we show that there are datasets where this upper bound on $\Delta p$ is non-trivial, and we therefore get even better utility.}

\subsection{Lipschitz Constant of the Cross-entropy Loss Function}
\label{sec:lip_cons}
In this section, we show how to calculate the upper bound on the Lipschitz constant $\rho$, which is important when computing the upper bound on global sensitivities $\Delta \omega$, $\Delta z_{T}$ and $\Delta p$ (see Equation~\eqref{eq:global_sensitivity_weight}).

\begin{lemma}~\cite{VirmauxA2018,GoukH2020}
\label{lm:lip_constant}
Given a fully connected neural network containing $T$ layers, $f_{T}: X \to \mathbb{R}^{C}$, where the hidden layers apply $1$-Lipschitz activation functions (e.g., ReLU, Tanh and Sigmoid), the output layer applies Softmax function, and $\mathbf{X}_{i}$ is the values of neurons at Layer $i$,  the Lipschitz constant (with respect to the model parameters) of $f_{T}$ is bounded by $\prod_{i=1}^{T} \lVert \mathbf{X}_{i} \rVert_{2}$.
\end{lemma}

\begin{lemma}~\cite{YedidaR2019}
\label{lm:one_layer_lip}
For a one-layer neural network with cross-entropy loss function, the Lipschitz constant of the cross-entropy loss function (with respect to the model parameters) is $\frac{C-1}{C|V|}\lVert \mathbf{X} \rVert_{2}$, where $C$ is the number of classes, $|V|$ is the number of the neurons at the input layer and $\mathbf{X}$ is a given data sample.
\end{lemma}

Based on Lemma~\ref{lm:lip_constant} and Lemma~\ref{lm:one_layer_lip}, we have Proposition~\ref{prop:lip_constant} to calculate the upper bound on the Lipschitz constant of the cross-entropy loss function, which is used in our experiments and also a commonly used loss function in existing works~\cite{ShokriR2017,SalemA2018,YeomS2018,JayaramanB2019}, for a $(T+1)$-layer neural network.

\begin{proposition}
\label{prop:lip_constant}
Using cross entropy function as the loss function of the aforementioned $(T+1)$-layer neural network, the upper bound on Lipschitz constant (with respect to the model parameters) is
\begin{align}
\label{eq:lip_constant}
    \rho \leq \frac{(C-1)\prod_{t=0}^{T-1}\sqrt{|V_{t}|}x_{t}}{C|V_{T-1}|},
\end{align}
where $C$ is the number of classes, $|V_{t}|$ is the number of neurons at layer $t$ and $x_{t}$ is the maximum absolute value of all the neurons at layer $t$.
\end{proposition}

\reviseA{Based on Equation~\eqref{eq:lip_constant}, the Lipschitz constant grows exponentially with the number of layers. However, when calculating the global sensitivities of the parameters $\Delta \omega$, $\Delta z_{T}$ and $\Delta p$, this exponential growth (in the number of layers and the maximum value of neuron per layer), is somewhat compensated by the terms $\sqrt{|W_{X}|}$ and $n$ (the number of data samples in the training set), as can be seen by putting the value of $\rho$ from Equation~\eqref{eq:lip_constant} into Equation~\eqref{eq:global_sensitivity_weight}. This is also empirically demonstrated in Table~\ref{tab:sensitivity} in Section~\ref{sec:exp_dp}. 
}

\subsection{Convexified Loss Function}
In our proofs of the upper bounds on the global sensitivities, a key assumption is that the objective function of the learning algorithm is a combination of a convex loss function and a $\lambda$-strongly convex regulariser. Lemma~\ref{lm:l2_norm_regulariser} guarantees the strong convexity for an $L_{2}$-norm regulariser. However, loss functions of deep learning models are non-convex due to a large number of model parameters~\cite{BengioY2006}. To circumvent this, we follow existing results in multi-layer (more than one hidden layers) neural networks convexification~\cite{LoJ2012,DvijothamK2014} by risk-averse optimisation to convexify the non-convex loss functions. 

\begin{theorem}~\cite{LoJ2012,DvijothamK2014}
\label{thm:convexification}
For a non-convex loss function $l(W_{X},\mathbf{x})$ and its training error $L_{X}(W_{X}) = \frac{1}{n}\sum_{i=1}^{n}l(W_{X},\mathbf{x}_{i})$, their risk-averting error functions, $l^{(\alpha)}(W_{X},\mathbf{x})$ and $L^{(\alpha)}_{X}(W_{X})$, are convex, where
\begin{align}
    & l^{(\alpha)}(W_{X},\mathbf{x}) = \exp(\alpha \times l(W_{X},\mathbf{x})), \nonumber \\
    & L^{(\alpha)}_{X}(W_{X}) = \frac{1}{\alpha}\ln\left[\frac{1}{n}\sum_{i=1}^{n}l^{(\alpha)}(W_{X},\mathbf{x}_{i})\right].
\end{align}
$\alpha$ is the risk-factor which measures the size of convex region. Larger $\alpha$ indicates larger convex region.
\end{theorem}

So we have
\begin{align}
\label{eq:convexification}
    & L^{(\alpha)}_{X}(W_{X}) + 2\lambda\lVert W_{X} \rVert^{2}_{2} \nonumber \\
    = & \frac{1}{\alpha}\ln\left[\frac{1}{n}\sum_{i=1}^{n}\exp(\alpha \times l(W_{X},\mathbf{x}_{i}))\right] + 2\lambda\lVert W_{X} \rVert^{2}_{2},
\end{align}
where $l(W_{X},\mathbf{x})$ can be conventional/commonly used non-convex loss functions, such as quadratic loss and cross-entropy loss. We use the convexified loss function rather than the traditional non-convex loss function in the experimental evaluations.

\reviseA{Intuitively, the convexification constant $\alpha$ might impact the trade-off between privacy and utility of the non-private model against MI attacks (since it might impact the performance/overfitting of the \revise{convexified} loss function~\cite{YeomS2018}). In this paper, we take $\alpha$ as another hyper-parameter to the machine learning model; hence, studying the relationship between $\alpha$ and the privacy-utility trade-off is left as future work.}

\subsection{Effect of Differential Privacy}
\label{sec:dp_effect}
In this section, we study the effect of differential privacy, including the DP guarantee of Algorithm~\ref{alg:dpdnn} and the \reviseA{OARO stability of a DP neural network.} \reviseA{In the $\epsilon$-DP proof of the algorithm, we assume the Laplace mechanism as an instance of $\text{Dist}(0, \sfrac{\Delta z_{T}}{\epsilon})$. But this guarantee can be converted into an $\epsilon$-GDP guarantee by invoking the advanced composition theorem of $(\epsilon, \delta(\epsilon))$-DP and subsequently applying Theorem~\ref{thm:gdp}. Thus, the result holds for both the Laplace and the Gaussian mechanisms.}

\begin{theorem}
\label{thm:final_dp}
The prediction probability vector $\mathbf{p}$ of a given observation $\mathbf{x}$, produced by Algorithm~\ref{alg:dpdnn} (injecting Laplace noise) is $(2C + 1)\epsilon$-differentially private, where $C$ is the number of classes.
\end{theorem}
\begin{proof}
Let the activation function at the output layer be Softmax and the values of the neurons prior to feeding Softmax be $\mathbf{z}_{T} = (z_{T,1}, \dots, z_{T,C})$, then we have the prediction probability vector $\mathbf{p} = (p_{1}, \dots, p_{C})$, where $p_{i} = \frac{\exp(z_{T,i})}{\sum_{i=1}^{C}\exp(z_{T,i})}$ and $C$ is the number of classes. 

Consider two sets of model parameters $W_{X}$ and $W_{X^{(-i)}}$ trained from two neighbouring dataset $X$ and $X^{(-i)}$, respectively. We run Algorithm~\ref{alg:dpdnn} on $W_{X}$ and $W_{X^{(-i)}}$ with the same privacy budget, the same data sample $\mathbf{x}$ for prediction and the same topology of the neural network. Based on Algorithm~\ref{alg:dpdnn}, we analyse the DP guarantee for the weighted sampling step and the noise injection step below.

Let the score function of the Exp-DP for sampling the neuron $v$ be $q(X,v) = p_{v}$. Based on Exp-DP and Lemma~\ref{cr:sensitivity_softmax}, the sampling weights of a neuron $v$ is $\Pr[\text{sample}(X,q,\epsilon) = v] = \frac{\exp(\epsilon q(X,v)/2\Delta q)}{\sum_{i=1}^{C}\exp(\epsilon q(X,i)/2\Delta q)}$, where $\Delta q = \exp(\Delta p)$. Following the standard proof of the Exp-DP~\cite{McsherryF2007}, we have
\begin{equation}
    \frac{\Pr[\text{sample}(X,q(X,v),\epsilon) = v]}{\Pr[\text{sample}(X^{(-i)},q(X^{(-i)},v),\epsilon) = v]} \leq \exp(\epsilon).
\end{equation}

For a neuron $v$ where noise has been injected and an arbitrary $r_{v} \in (0,1)$, we have the worst-case upper bound on the probability ratio for $p_{u} \in \mathbf{p}$ ($\mathbf{p}$ is the noisy prediction probability vector) as follows.
\begin{align}
\label{eq:noisy_neuron}
    & \frac{\Pr[\hat{p}_{v} = r_{v} | \text{sample}(X,q(X,v),\epsilon) = v]}{\Pr[\hat{p}_{v}^{(-i)} = r_{v} | \text{sample}(X^{(-i)},q(X^{(-i)},v),\epsilon) = v]} \nonumber \\
    = & \frac{\Pr\left[\frac{\exp(\hat{z}_{T,v})}{\exp(\hat{z}_{T,v}) + \sum_{j \neq v}\exp(z_{T,j})} = r_{v}\right]}{\Pr\left[\frac{\exp(\hat{z}_{T,v}^{(-i)})}{\exp(\hat{z}_{T,v}^{(-i)}) + \sum_{j \neq v}\exp(z_{T,j}^{(-i)})} = r_{v}\right]} \nonumber \\
    = & \frac{\Pr\left[\hat{z}_{T,v} = \ln\left(\frac{r_{v}}{1-r_{v}}\sum_{j \neq v}\exp(z_{T,j})\right)\right]}{\Pr\left[\hat{z}_{T,v}^{(-i)} = \ln\left(\frac{r_{v}}{1-r_{v}}\sum_{j \neq v}\exp(z_{T,j}^{(-i)})\right)\right]} \nonumber \\
    = & \frac{\exp\left(\frac{\epsilon}{\Delta z_{T}}\left|\ln\left(\frac{r_{v}}{1-r_{v}}\sum_{j \neq v}\exp(z_{T,j})\right) - z_{T,v}\right|\right)}{\exp\left(\frac{\epsilon}{\Delta z_{T}}\left|\ln\left(\frac{r_{v}}{1-r_{v}}\sum_{j \neq v}\exp(z_{T,j}^{(-i)})\right) - z_{T,v}^{(-i)}\right|\right)} \nonumber \\
    \leq & \exp\left(\frac{\epsilon}{\Delta z_{T}}\left|\ln\left(\frac{\sum_{j \neq v}\exp(z_{T,j}))}{\sum_{j \neq v}\exp(z_{T,j}^{(-i)}))}\right) - z_{T,v} + z_{T,v}^{(-i)}\right|\right) \nonumber \\
    \leq & \exp\left(\frac{\epsilon}{\Delta z_{T}}\left|\ln\left(\frac{\sum_{j \neq v}\exp(z_{T,j}^{(-i)} + \Delta z_{T}))}{\sum_{j \neq v}\exp(z_{T,j}^{(-i)}))}\right) + \Delta z_{T}\right|\right) \nonumber \\
    = & \exp\left(\frac{\epsilon}{\Delta z_{T}}\left|\ln\left(\exp(\Delta z_{T})\frac{\sum_{j \neq v}\exp(z_{T,j}^{(-i)}))}{\sum_{j \neq v}\exp(z_{T,j}^{(-i)}))}\right) + \Delta z_{T}\right|\right) \nonumber \\
    = & \exp\left(\frac{\epsilon}{\Delta z_{T}}\big|\Delta z_{T} + \Delta z_{T}\big|\right) \nonumber \\
    = & \exp(2\epsilon).
\end{align}

For an arbitrary neuron $u$ where noise has not been injected (we inject noise into a neuron $v$) and an arbitrary $r_{u} \in (0,1)$, we have the worst-case upper bound on the probability ratio for $\hat{p}_{u} \in \mathbf{p}$ ($\mathbf{p}$ is the noisy prediction probability vector) under the condition of $\hat{p}_{v} = r_{v}$ and $\hat{p}_{v}^{(-i)} = r_{v}$.
\begin{align}
    & \frac{\Pr[\hat{p}_{u} = r_{u} | \hat{p}_{v} = r_{v}]}{\Pr[\hat{p}_{u}^{(-i)} = r_{u} | \hat{p}_{v}^{(-i)} = r_{v}]} \nonumber \\
    = & \frac{\Pr\left[\frac{\exp(z_{T,u})}{\sum_{j=1}^{C}\exp(z_{T,j})} = r_{u} \Big| \frac{\exp(\hat{z}_{T,v})}{\sum_{j=1}^{C}\exp(z_{T,j})} = r_{v}\right]}{\Pr\left[\frac{\exp(z_{T,u}^{(-i)})}{\sum_{j=1}^{C}\exp(z_{T,j}^{(-i)})} = r_{u} \bigg| \frac{\exp(\hat{z}_{T,v}^{(-i)})}{\sum_{j=1}^{C}\exp(z_{T,j}^{(-i)})} = r_{v}\right]} \nonumber \\
    = & \frac{\Pr\left[r_{v} \cdot \frac{\exp(z_{T,u})}{\exp(\hat{z}_{T,v})} = r_{u}\right]}{\Pr\left[r_{v} \cdot \frac{\exp(z_{T,u}^{(-i)})}{\exp(\hat{z}_{T,v}^{(-i)})} = r_{u}\right]} \nonumber \\
    = & \frac{\Pr\left[\hat{z}_{T,v} = \ln\left(\frac{r_{v}\exp(z_{T,u})}{r_{u}}\right)\right]}{\Pr\left[\hat{z}_{T,v}^{(-i)} = \ln\left(\frac{r_{v}\exp(z_{T,u}^{(-i)})}{r_{u}}\right)\right]} \nonumber \\
    = & \frac{\exp\left(\frac{\epsilon}{\Delta z_{T}}\left|\ln\left(\frac{r_{v}\exp(z_{T,u})}{r_{u}}\right) - z_{T,v}\right|\right)}{\exp\left(\frac{\epsilon}{\Delta z_{T}}\left|\ln\left(\frac{r_{v}\exp(z_{T,u}^{(-i)})}{r_{u}}\right) - z_{T,v}^{(-i)}\right|\right)} \nonumber \\
    \leq & \exp\left(\frac{\epsilon}{\Delta z_{T}}\left|\ln\left(\frac{\exp(z_{T,u})}{\exp(z_{T,u}^{(-i)})}\right) - z_{T,v} + z_{T,v}^{(-i)}\right|\right) \nonumber \\
    \leq & \exp\left(\frac{\epsilon}{\Delta z_{T}}\left|\ln\left(\frac{\exp(z_{T,u}^{(-i)} + \Delta z_{T})}{\exp(z_{T,u}^{(-i)})}\right) + \Delta z_{T}\right|\right) \nonumber \\
    = & \exp\left(\frac{\epsilon}{\Delta z_{T}}\left|\ln\left(\exp(\Delta z_{T})\frac{\exp(z_{T,u}^{(-i)}))}{\exp(z_{T,u}^{(-i)}))}\right) + \Delta z_{T}\right|\right) \nonumber \\
    = & \exp\left(\frac{\epsilon}{\Delta z_{T}}\big|\Delta z_{T} + \Delta z_{T}\big|\right) \nonumber \\
    = & \exp(2\epsilon).
\end{align}

Then we give the upper bound on the privacy guarantee for the final prediction result.
\begin{align}
\label{eq:final_dp}
    & \frac{\Pr[\hat{\mathbf{p}} = \mathbf{r}]}{\Pr[\hat{\mathbf{p}}^{(-i)} = \mathbf{r}]} \nonumber \\
    = & \frac{\Pr[\text{sample}(X,\mathbf{p},\epsilon) = v]}{\Pr[\text{sample}(X^{(-i)},\mathbf{p}^{(-i)},\epsilon) = v]} \times \prod_{j=1}^{C}\frac{\Pr[\hat{p}_{j} = r_{j}]}{\Pr[\hat{p}^{(-i)}_{j} = r_{j}]} \nonumber \\
    \leq & \exp(\epsilon) \times \frac{\Pr[\hat{p}_{v} = r_{v}]}{\Pr[\hat{p}^{(-i)}_{v} = r_{v}]} \times \prod_{j \neq v}\frac{\Pr[\hat{p}_{j} = r_{j} | \hat{p}_{v} = r_{v}]}{\Pr[\hat{p}^{(-i)}_{j} = r_{j} | \hat{p}^{(-i)}_{v} = r_{v}]} \nonumber \\
    \leq & \exp(\epsilon) \times \prod_{j=1}^{C}\exp(2\epsilon) \nonumber \\
    = & \exp((2C+1)\epsilon).
\end{align}

This concludes the proof.
\end{proof}

\begin{corollary}
\label{cr:dp_effect}
\reviseA{A differentially private neural network is more On-Average-Remove-One stable than a non-private neural network.}
\end{corollary}
\begin{proof}
Injecting DP noise, either by objective perturbation~\cite{PhanN2016,PhanN2017} or gradient perturbation (as DP-SGD~\cite{AbadiM2016}) or output perturbation (as Algorithm~\ref{alg:dpdnn}), could be modelled as modifying the model parameters to be $\hat{W}_{X}$. Note that $\hat{W}_{X}$ is not a minimiser of $L_{X}(W_{X}) + \frac{\lambda}{2}\lVert W_{X} \rVert^{2}_{2}$, replacing $W_{X}$ in Inequality~\eqref{subeq:lambda_strong} by $\hat{W}_{X}$ decreases the value of Inequality~\eqref{subeq:lambda_strong} and Equation~\eqref{subeq:change_measure}. Following the same proof as Theorem~\ref{thm:global_sensitivity}, we have a tight upper bound for $\lVert W_{X^{(-i)}} - \hat{W}_{X} \rVert_{2}$, say, with a $c > 0$, $\lVert W_{X^{(-i)}} - \hat{W}_{X} \rVert_{2} = \lVert W_{X^{(-i)}} - W_{X} \rVert_{2} - c \leq \sfrac{2\rho}{\lambda n} - c$. Since we assume the loss function is $\rho$-Lipschitz, we have
\begin{align}
    l(W_{X^{(-i)}},\mathbf{x}_{i}) - l(\hat{W}_{X},\mathbf{x}_{i}) & \leq \rho\lVert W_{X^{(-i)}} - \hat{W}_{X} \rVert_{2} \nonumber \\
    & = \rho\lVert W_{X^{(-i)}} - W_{X} \rVert_{2} - c\rho \nonumber \\
    & \leq \frac{2\rho^{2}}{\lambda n} - c\rho.
\end{align}
Since this is valid for any $X$ and $\mathbf{x}_{i}$, $i \sim U(|X|)$, $|X| = n$, we immediately have
\begin{equation}
\mathbb{E}_{(X, \mathbf{x}_{i}) \sim \mathcal{D}^{n+1}, i \sim U(n)}[l(W_{X^{(-i)}}, \mathbf{x}_{i}) - l(\hat{W}_{X}, \mathbf{x}_{i})] \leq \frac{2\rho^{2}}{\lambda n} - c\rho.
\end{equation}
Since $c\rho > 0$ and the upper bound on $\mathbb{E}_{(X, \mathbf{x}_{i}) \sim \mathcal{D}^{n+1}, i \sim U(n)}[l(W_{X^{(-i)}}, \mathbf{x}_{i}) - l(W_{X}, \mathbf{x}_{i})]$ is $\sfrac{2\rho^{2}}{\lambda n}$ (Theorem~\ref{thm:global_sensitivity}), injecting DP noise makes the model more On-Average-Remove-One stable. 
\end{proof}

\reviseA{Based on Definition~\ref{def:remove_stable}, the more On-Average-Remove-One stable a model, the less likely it is to overfit. As shown in \cite{YeomS2018}, a less overfitted (or more generalised) model is more resistant against MI attacks.}

\section{Experimental Evaluation}
\label{sec:exp}
In this section, we show experimental evaluations of the proposed algorithm. We start with a description of the datasets, the performance metrics, experimental configurations and finally evaluation results and analyses. \reviseA{We are providing an open-source implementation of our algorithm to aid future research.\footnote{See \url{https://github.com/suluz/dp_ml_API}}}

\subsection{Datasets}
In the experiments, we use the same datasets as Shokri et al.~\cite{ShokriR2017} and Jayaraman et al.~\cite{JayaramanB2019}, i.e., US Adult (Income)\footnote{\url{http://archive.ics.uci.edu/ml/datasets/Adult}}, MNIST\footnote{\url{http://yann.lecun.com/exdb/mnist/}}, Location (Bangkok restaurants check-ins)\footnote{\url{https://sites.google.com/site/yangdingqi/home/foursquare-dataset}}, Purchases\footnote{\url{https://www.kaggle.com/c/acquire-valued-shoppers-challenge/data}}, CIFAR\footnote{\url{https://www.cs.toronto.edu/~kriz/cifar.html}} and Texas Hospital\footnote{\url{https://www.dshs.texas.gov/THCIC/Hospitals/Download.shtm}}. Since these datasets are commonly used in the field of machine learning and MI attacks, we only show the statistics of them in Table~\ref{tab:datasets}, where \#Rec. is the number of records (randomly sampled from the raw datasets) and \#Feat. is the number of features/attributes in training sets. For the details of these datasets, please refer to Section IV-a of \cite{ShokriR2017} and Section 4.1 of \cite{JayaramanB2019}. 
\begin{table}[!ht]
\centering
\caption{Datasets Statistics.}
\label{tab:datasets}
\scalebox{0.9}{
\begin{tabular}{l|c|c|c|c}
\hline
Dataset & \#Rec. & \#Feat. & \#Classes & \#Shadow Models\\
\hline\hline
US Adult & 10,000 & 14 & 2 & 20 \\ 
\hline
MNIST & 10,000 & 784 & 10 & 50 \\ 
\hline
Location & 1,200 & 446 & 30 & 30 \\ 
\hline
Purchase-2 & 10,000 & 600 & 2 & 20 \\ 
\hline
Purchase-10 & 10,000 & 600 & 10 & 20 \\
\hline
Purchase-20 & 10,000 & 600 & 20 & 20 \\
\hline
Purchase-50 & 10,000 & 600 & 50 & 20 \\
\hline
Purchase-100 & 10,000 & 600 & 100 & 20 \\
\hline
CIFAR-10 & 10,000 & 3,072 & 10 & 100 \\ 
\hline
CIFAR-100 & 10,000 & 3,072 & 100 & 100 \\ 
\hline
Texas Hospital & 10,000 & 6,169 & 100 & 10 \\ 
\hline
\end{tabular}
}
\end{table}

\subsection{Performance Metrics}
\descr{Metrics for overfitting of the baseline non-private models.} In our experiments, we use the upper bound on the On-Average-Remove-One (OARO) Stability (Definition~\ref{def:remove_stable} and Equation~\eqref{eq:upper_loss} in Theorem~\ref{thm:global_sensitivity}) to measure the overfitting of baseline non-private models, that is, 
\begin{equation}
\label{eq:oaro}
    \mathbb{E}_{(X, \mathbf{x}_{i}) \sim \mathcal{D}^{n+1}, i \sim U(n)}[l(W_{X^{(-i)}}, \mathbf{x}_{i}) - l(W_{X}, \mathbf{x}_{i})] \leq \frac{2\rho^{2}}{\lambda n},
\end{equation}
where $\rho$ is the maximum Lipschitz constant on a given training set and a given neural networks, $\lambda$ is the strongly convex constant shown in Lemma~\ref{lm:l2_norm_regulariser} and $n$ is the size of the training set. \reviseA{In practice, for a given training set and a given neural network, $C$, $T$ and $|V_{t}|$ ($t \in [0, T]$) are fixed and pre-determined. We further take the maximum value of $x_{t}$ at each layer to calculate the upper bound on Lipschitz constant $\rho$ by Equation~\eqref{eq:lip_constant} in Proposition~\ref{prop:lip_constant}. We then use this empirical maximum $\rho$ to calculate the upper bound on the global sensitivities.} Note that, a tight upper bound on Inequality~\eqref{eq:oaro} indicates more OARO stable.

\descr{Metrics for DP models.} Following existing studies~\cite{YeomS2018,JayaramanB2019} on measuring DP performance against MI attacks, we use the same metrics in this paper, i.e., accuracy loss, a DP model's accuracy loss on the test set with respect to the baseline non-private model and privacy leakage, the difference between the true positive rate and the false positive rate of the MI attacks (as binary classifiers). They are defined as follows.
\begin{align}
    & \text{Acc\_Loss} = 1 - \frac{\text{Test\_Acc}_{\text{DP}}}{\text{Test\_Acc}_{\text{Baseline (S)}}}, \label{subeq:acc_loss}\\
    & \text{Priv\_Leak} = \frac{|\text{TP}_{\text{MI}}|}{|\text{P}|} - \frac{|\text{FP}_{\text{MI}}|}{|\text{N}|}, \label{subeq:priv_leak}
\end{align}
where Baseline (S) indicates the baseline non-private model trained on the \underline{S}urrogate loss function, $\text{TP}_{\text{MI}}$ and $\text{FP}_{\text{MI}}$ are the true positive and false positive of the MI attacks (specifying as the black-box shadow model-based approach from Shokri et al.~\cite{ShokriR2017} in this paper), P and N are positive/member and negative/non-member labels of a test data sample. \reviseA{Note that, based on Equation~\eqref{subeq:acc_loss} and Equation~\eqref{subeq:priv_leak} both $\text{Acc\_Loss}$ and $\text{Priv\_Leak}$ are in $[0, 1]$. $\text{Acc\_Loss} = 0$ indicates that a DP model achieved the maximum prediction accuracy, i.e., the same as the baseline model and $\text{Acc\_Loss} = 1$ indicates no prediction utility of a DP model. $\text{Priv\_Leak} = 0$ indicates no privacy leakage of a given model and $\text{Priv\_Leak} = 1$ indicates that a given model leaks the maximum privacy under MI attacks.}

\subsection{Experimental Configurations}
Since existing upper bounds on the global sensitivity of the trained model parameters are looser than ours under the same condition -- convexity of the loss function -- our upper bound is expected to outperform existing upper bounds given the same output perturbation-based algorithm. Hence we do not perform experimental comparison between our upper bound and existing upper bounds. In our experiments, we compare the performance achieved by Algorithm~\ref{alg:dpdnn} and Google's open-source implementation~\cite{TensorflowPrivacy} of DP-SGD approaches on the real-world datasets.

\begin{table*}[!ht]
\centering
\caption{Configurations in the Experiments.}
\label{tab:configurations}
\scalebox{0.9}{
\begin{tabular}{c|c|c|c|c|c|c}
\hline
\multicolumn{7}{c}{Non-Private/DP-SGD/MI Shadow Models} \\ \hline
\#hidden neurons   & $L_{2}$ reg.  & learning rate    & \#hidden layer  & optimiser & batch size   & \#epoch \\ \hline
\textit{128} & \textit{0.001} & \textit{0.001} & \textit{1} & \textit{ADMA}  & \textit{100}   & \textit{100} \\ \hline
activation func.    & loss func.    & $\epsilon$    & $\delta$  & \multicolumn{3}{c}{DP Implementation} \\ \hline
\textit{Tanh}, \textit{Softmax} & \textit{cross-entropy}   & \textit{[0.01, 1000]}    & $\sfrac{1}{10 \times |X|}$ & \multicolumn{3}{c}{\textit{RDP}, \textit{zCDP}, \textit{AC}, \textit{NC}} \\ \hline
\multicolumn{7}{c}{Convexification Constant ($\alpha$ in Equation~\eqref{eq:convexification})} \\ \hline
\multicolumn{5}{c|}{$\alpha = 1$: Location, MNIST, US Adult, Purchase-2, CIFAR-100} & \multicolumn{2}{c}{$\alpha = 2$: CIFAR-10} \\ \hline
\multicolumn{2}{c|}{$\alpha = 4$: Texas Hospital} & \multicolumn{5}{c}{$\alpha = 5$: Purchase-100, Purchase-50, Purchase-20, Purchase-10} \\ \hline\hline
\multicolumn{7}{c}{MI Attack Models} \\ \hline
\#hidden neurons   & $L_{2}$ reg.  & learning rate  &   \multicolumn{4}{c}{Other Configurations} \\ \hline
\textit{64}  & \textit{$1e^{-6}$}  & \textit{0.01}   & \multicolumn{4}{c}{\textit{The Same as MI Shadow Models}} \\ \hline
\end{tabular}
}
\end{table*}
\descr{Hyper-parameters.} In general, we mainly follow the configurations in existing studies on MI attacks (the shadow model-based approach from Shokri et al.)~\cite{ShokriR2017} and DP-SGD~\cite{JayaramanB2019} but train baseline non-private models, MI shadow models and DP models with a convexified loss function. Specifically, for training baseline non-private models and MI shadow models~\cite{ShokriR2017}, we keep the same configurations as Shokri et al.~\cite{ShokriR2017}; to implement the DP-SGD (injecting Gaussian noise) approaches, we call Google's open-source tensorflow-privacy package following the same configurations as Jayaraman et al.~\cite{JayaramanB2019}. We also follow the same computation as Jayaraman et al.~\cite{JayaramanB2019} to plot the theoretical upper bound on the privacy leakage of $\epsilon$-DP, which is $\exp(\epsilon) - 1$. When plotting it, we bound it by the privacy leakage of the baseline non-private model. We empirically search the value of convexification constant ($\alpha$ in Equation~\eqref{eq:convexification}) \reviseA{in $(0, 10]$ with a step of $0.5$} to ensure the \revise{convexified} loss function achieves (almost) the same training \reviseA{and test accuracy} as the original (non-convex) loss function. Table~\ref{tab:configurations} shows the detailed configurations, where \textit{RDP}, \textit{zCDP}, \textit{AC} and \textit{NC} represent R{\'e}nyi DP~\cite{MironovI2017}, zero-Concentrated DP~\cite{BunM2016}, DP with advanced composition~\cite{DworkC2014} and DP with na{\"i}ve composition~\cite{DworkC2006O}, respectively, $|X|$ is the size of the training set.

\descr{Privacy budget.} \reviseA{In our experiments, we report the average performance of all DP models with the same privacy budget $\epsilon$ answering a single query (the prediction vector of a single data sample). This is mostly for ease of presentation. See Section~\ref{sub:budget-multiple-queries} for a discussion of how this can be expanded to multiple queries.} We apply both Laplace and Gaussian mechanisms to generate DP noise for Algorithm~\ref{alg:dpdnn}. In particular, when applying Laplace distribution, according to the sequential composition of privacy budget (Theorem~\ref{thm:Lap-DP_composition}), we split the overall privacy budget $\epsilon$ for a single query as Theorem~\ref{thm:final_dp}, that is, $\epsilon_{\text{neuron}} (= \epsilon_{\text{sampling}}) = \sfrac{\epsilon}{(2C + 1)}$. When applying Gaussian distribution, according to the relationship between $\epsilon$-GDP and $(\epsilon, \delta)$-DP (Theorem~\ref{thm:GDP_composition}) and the composition theorem proposed in Gaussian differential privacy (Theorem~\ref{thm:GDP_composition}), we split the overall privacy budget $\epsilon$ as $\epsilon_{\text{neuron}} (= \epsilon_{\text{sampling}}) = \sfrac{\epsilon}{\sqrt{2^{2} \times C + 1}}$. \reviseA{When implementing DP-SGD, we use the source code of Jayaraman et al.~\cite{JayaramanB2019} to split $\epsilon$ into each epoch.} The privacy budget values in this paper are the same as Jayaraman et al.~\cite{JayaramanB2019}, i.e., $\epsilon = \{0.01, 0.05, 0.1, 0.5, 1, 5, 10, 50, 100, 500, 1000\}$. 
 
\descr{Experimental setup.} To evaluate the DP models (Algorithm~\ref{alg:dpdnn} and DP-SGD models) on a given dataset with a privacy budget $\epsilon$ \reviseA{for a single data sample for prediction}, we first randomly sample two disjoint sets having the same size from the dataset to be the training set and the test set. Using the training set, we train the baseline non-private model and the $\epsilon$-DP-SGD models. We then implement Algorithm~\ref{alg:dpdnn} with $\epsilon$ on the test set. We calculate the accuracy loss of each DP model based on the test accuracy obtained from the test set. Then we perform the MI attack via shadow models~\cite{ShokriR2017} to attack the baseline non-private model and the DP models to calculate the privacy leakage. Finally, we repeat the training and attacking process 10 times to report the average accuracy loss and privacy leakage. 

\subsection{Experimental Results and Analysis}
In this section, we empirically study the non-private model and DP models on real-world datasets. We first report the performance, including training and test accuracy, OARO stability and privacy leakage, of the baseline non-private model using the \revise{convexified} loss function. Then we present the comparison between Algorithm~\ref{alg:dpdnn} and DP-SGD on accuracy loss and privacy leakage. Finally we analyse the experimental results (accuracy loss and privacy leakage) of the DP models (Algorithm~\ref{alg:dpdnn} and existing DP-SGD) obtained on the real-world datasets.

\subsubsection{Performance of the Convexified Loss Function}
\label{sec:exp_surrogate}
We compare the performance of models trained on the original non-convex loss function and on the convexified loss function with the same hyper-parameters and the same training and training sets. Table~\ref{tab:accuracy} shows the average training accuracy and the average test accuracy achieved on the real-world datasets of $10$ experiments.

\begin{table*}[!ht]
\centering
\caption{Performance of the Baseline Non-Private Models\\(sorted in decreasing order of Acc. Gap/Priv. Leak. of overfitting and fitting models).}
\label{tab:accuracy}
\scalebox{0.9}{
\begin{tabular}{l|c|c|c|c|c|c|c|c}
\hline
\multirow{2}{*}{Dataset} & \multicolumn{3}{c|}{Original Loss Func.} & \multicolumn{3}{c|}{\revise{Convexified} Loss Func.} & \multirow{2}{*}{Priv. Leak.} & \multirow{2}{*}{OARO Stability} \\ \cline{2-7}
& Training Acc. & Test Acc. & Acc. Gap & Training Acc. & Test Acc. & Acc. Gap &  \\
\hline\hline
Location & 1.0000 & 0.6068 & 0.3932 & 0.9916 & 0.6484 & 0.3432 & 0.3600 & 10.8532 \\ 
\hline
Texas Hospital & 0.7990 & 0.5689 & 0.2301 & 0.8770 & 0.5364 & 0.3406 & 0.2295 & 18.8945 \\ 
\hline
Purchase-100 & 0.9992 & 0.7985 & 0.2007 & 0.9942 & 0.7723 & 0.2219 & 0.1664 & 1.8193 \\
\hline
Purchase-50 & 0.9994 & 0.8636 & 0.1358 & 0.9985 & 0.8315 & 0.1670 & 0.1236 & 1.7827 \\
\hline\hline
Purchase-20 & 0.9986 & 0.9022 & 0.0964 & 0.9796 & 0.8759 & 0.1037 & 0.0727 & 1.6750 \\
\hline
Purchase-10 & 0.9973 & 0.9203 & 0.0770 & 0.9735 & 0.8903 & 0.0832 & 0.0292 & 1.5036 \\
\hline
Purchase-2 & 0.9963 & 0.9642 & 0.0321 & 0.9951 & 0.9642 & 0.0309 & 0.0073 & 0.4641 \\ 
\hline
MNIST & 0.9863 & 0.9528 & 0.0335 & 0.9494 & 0.9297 & 0.0197 & 0.0035 & 1.9845 \\ 
\hline
US Adult & 0.8310 & 0.8300 & 0.0010 & 0.8260 & 0.8262 & 0.0002 & 0.0023 & 0.0109 \\ 
\hline\hline
CIFAR-10 & 0.6198 & 0.4453 & 0.1745 & 0.5731 & 0.4236 & 0.1495 & 0.0111 & 7.7760 \\ 
\hline
CIFAR-100 & 0.3224 & 0.1677 & 0.1647 & 0.2026 & 0.1389 & 0.0637 & 0.0099 & 9.4090 \\ 
\hline
\end{tabular}
}
\end{table*}

As shown in Table~\ref{tab:accuracy}, the \revise{convexified} loss function provides approximately the same model performance (training accuracy and test accuracy) as the original non-convex loss function. Such a result is also confirmed by the experimental analysis in \cite{DvijothamK2014}. Therefore, we can use the \revise{convexified} loss function to provide reliable baseline non-private models to further analyse the performance of the DP models. 


\subsubsection{Performance of the On-Average-Remove-One Stability on Measuring Overfitting}
\label{sec:exp_stability}
In this section, we examine the performance of the OARO stability on measuring overfitting via checking the relationship between the OARO stability and two empirical rules of detecting overfitting~\cite{YeomS2018}, i.e., accuracy gap between the training accuracy and the test accuracy and privacy leakage under the MI attacks. 

\begin{figure}[!ht]
    \centering
    \captionsetup{justification=centering}
    \subfloat[OARO Stability \\vs. Acc. Gap]{
    \includegraphics[width=0.15\textwidth]{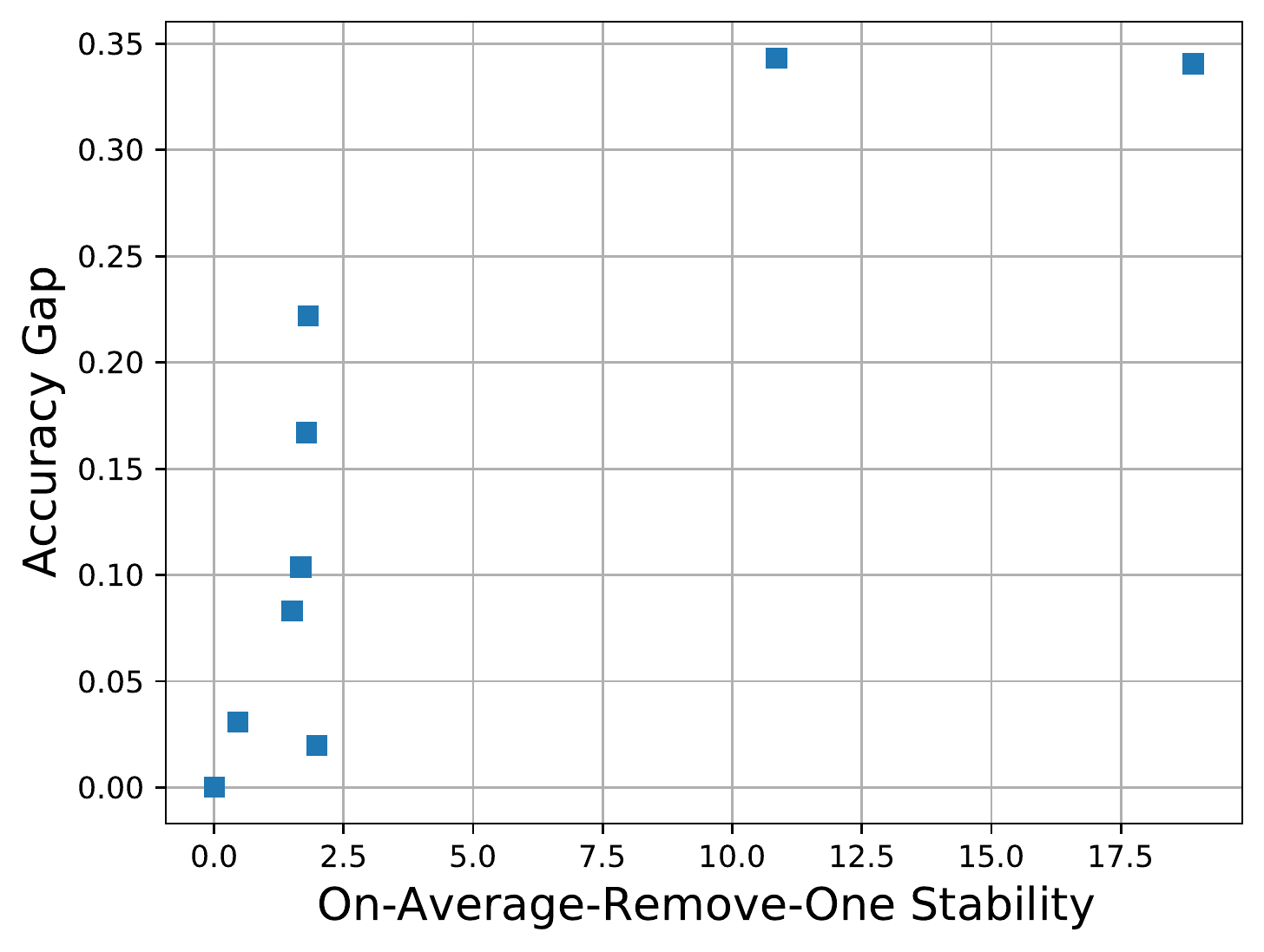}
    \label{subfig:acc_gap_stability}
    }
    \hfill
    \subfloat[OARO Stability \\vs. Priv. Leak.]{
    \includegraphics[width=0.15\textwidth]{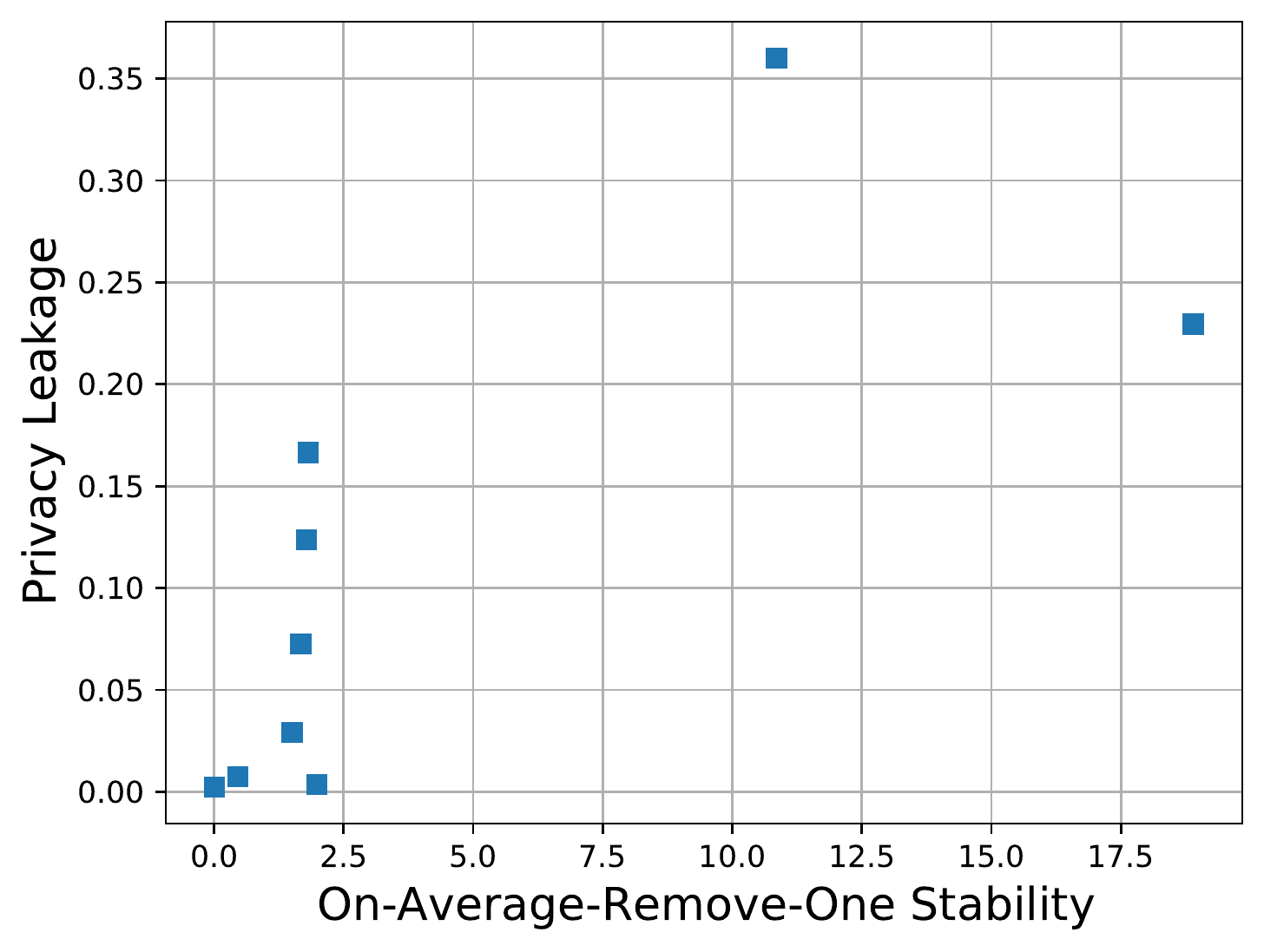}
    \label{subfig:priv_leak_stability}
    }
    \hfill
    \subfloat[Priv. Leak. \\vs. Acc. Gap]{
    \includegraphics[width=0.15\textwidth]{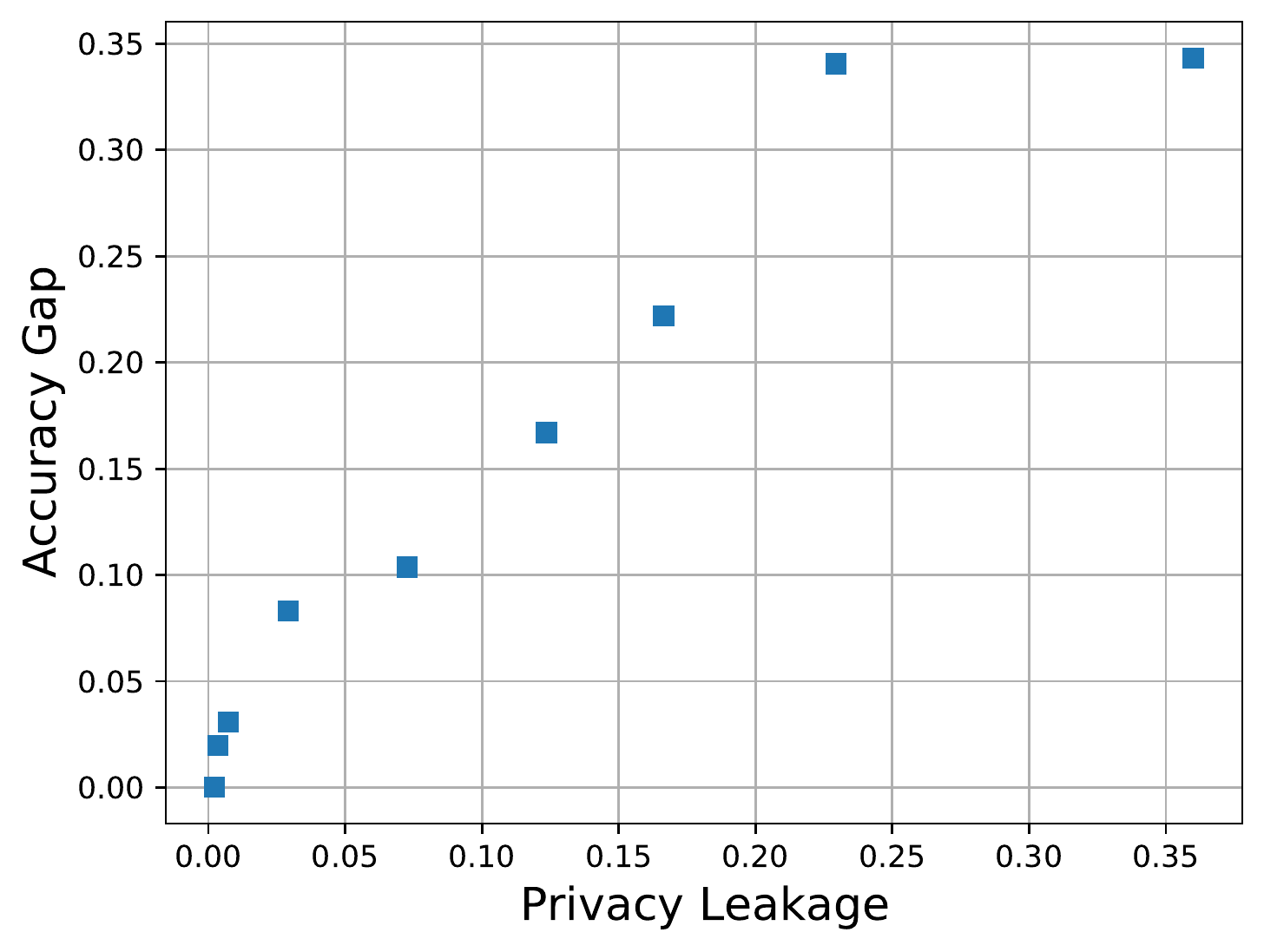}
    \label{subfig:priv_leak_acc_gap}
    }
    \caption{OARO Stability vs. Empirical Rules of Overfitting.}
    \label{fig:stability}
\end{figure}

Table~\ref{tab:accuracy} and Figure~\ref{fig:stability} show that the OARO stability is significantly correlated to both the accuracy gap and the privacy leakage, when the training accuracy of a baseline non-private model is acceptable (test accuracy is greater than $0.5$). Specifically, in Figure~\ref{subfig:acc_gap_stability}, the Pearson Correlation Coefficient is 0.8277 (p-value = 0.0059). In Figure~\ref{subfig:priv_leak_stability}, the Pearson Correlation Coefficient is 0.7412 (p-value = 0.0223). In both Figures~\ref{subfig:acc_gap_stability} and \ref{subfig:priv_leak_stability}, the Spearman's rank Correlation Coefficient is 0.7333 (p-value = 0.0246), the Kendall's rank Correlation Coefficient is 0.6667 (p-value = 0.0127). Therefore, we conclude that \textbf{the OARO stability can estimate the overfitting of a model when the model has a high training accuracy,} since we calculate the OARO stability during the training process. This could be a potential benefit to detect overfitting of a model when there are not enough data to prepare training sets to compare the accuracy gap.

\subsubsection{Performance of the Differentially Private Models}
\label{sec:exp_dp}
\begin{figure}[!ht]
    \centering
    \captionsetup{justification=centering}
    \subfloat[Accuracy Loss]{
    \includegraphics[width=0.23\textwidth]{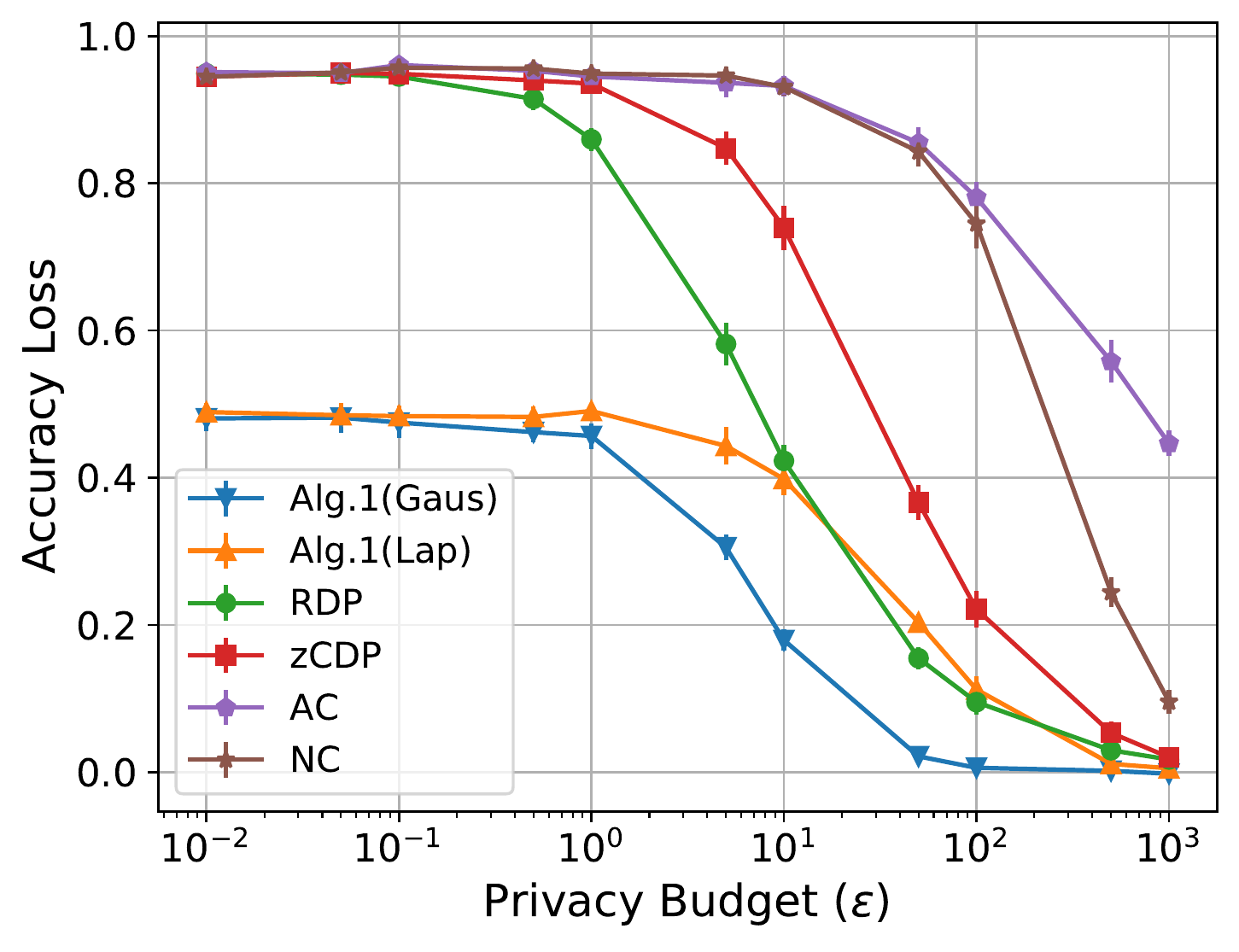}
    \label{subfig:location_acc_loss}
    }
    \hfill
    \subfloat[Privacy Leakage]{
    \includegraphics[width=0.23\textwidth]{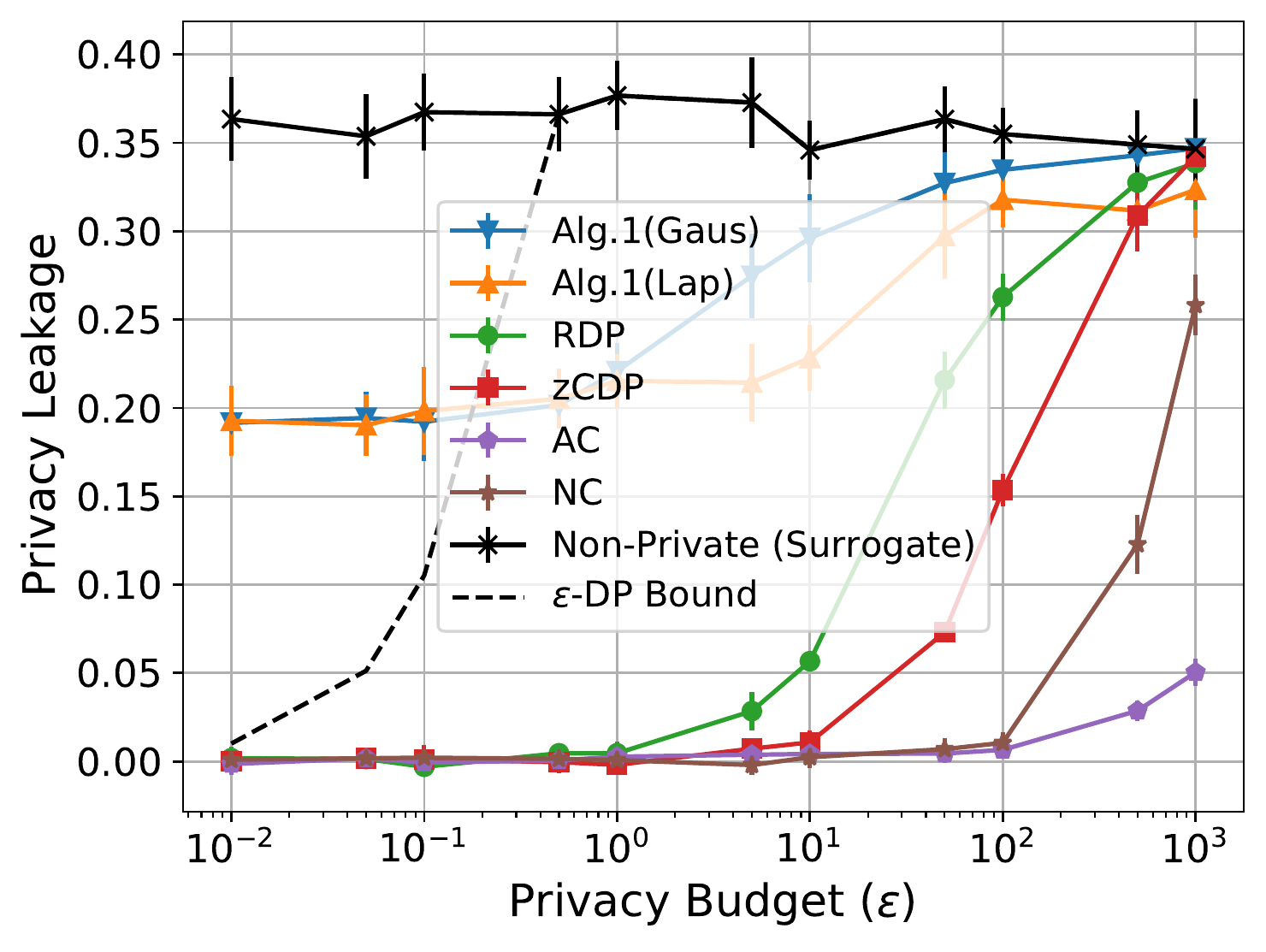}
    \label{subfig:location_priv_leak}
    }
    \caption{Performance Evaluations on Location Dataset.}
    \label{fig:location}
\end{figure}

\begin{figure}[!ht]
    \centering
    \captionsetup{justification=centering}
    \subfloat[Accuracy Loss]{
    \includegraphics[width=0.23\textwidth]{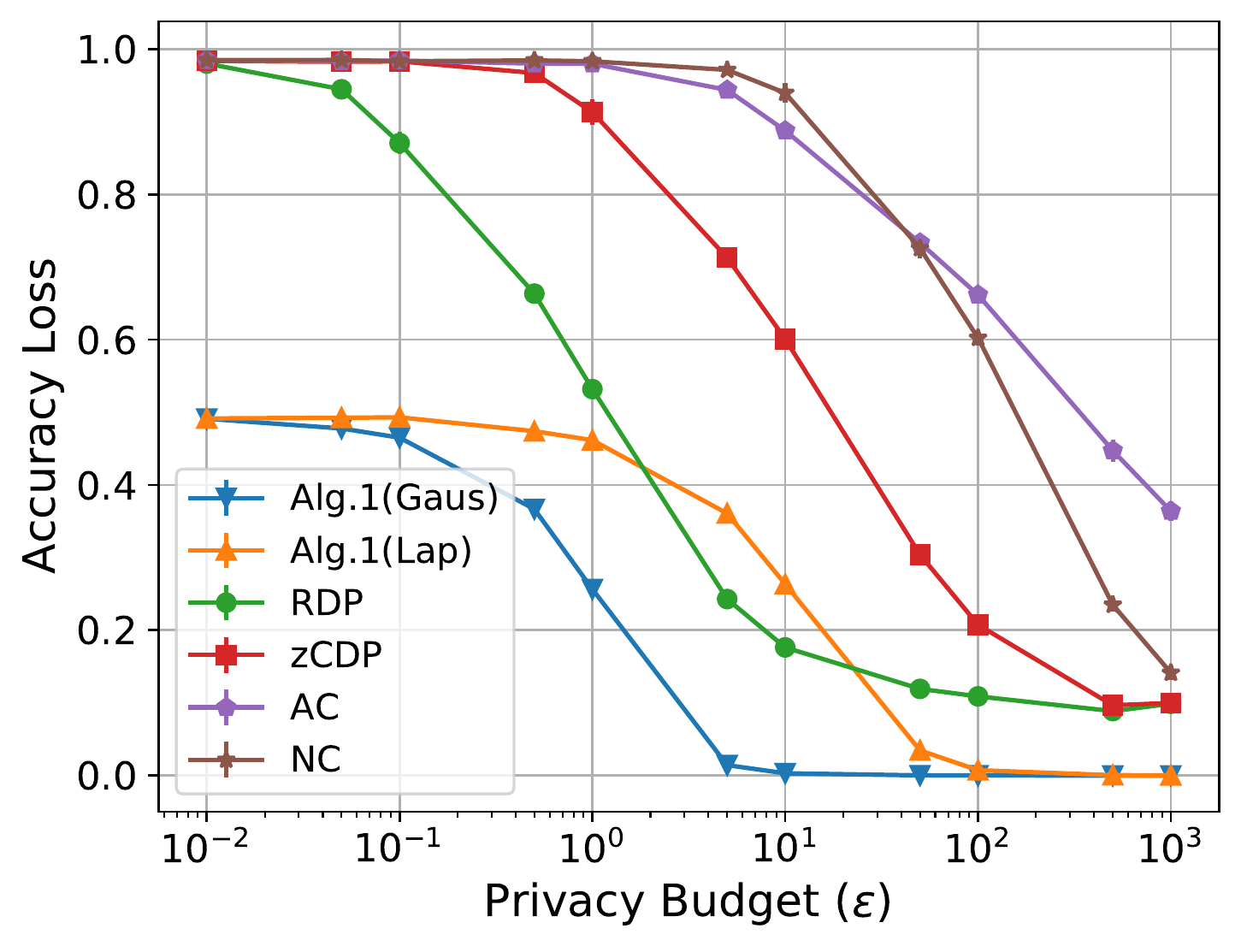}
    \label{subfig:texas_acc_loss}
    }
    \hfill
    \subfloat[Privacy Leakage]{
    \includegraphics[width=0.23\textwidth]{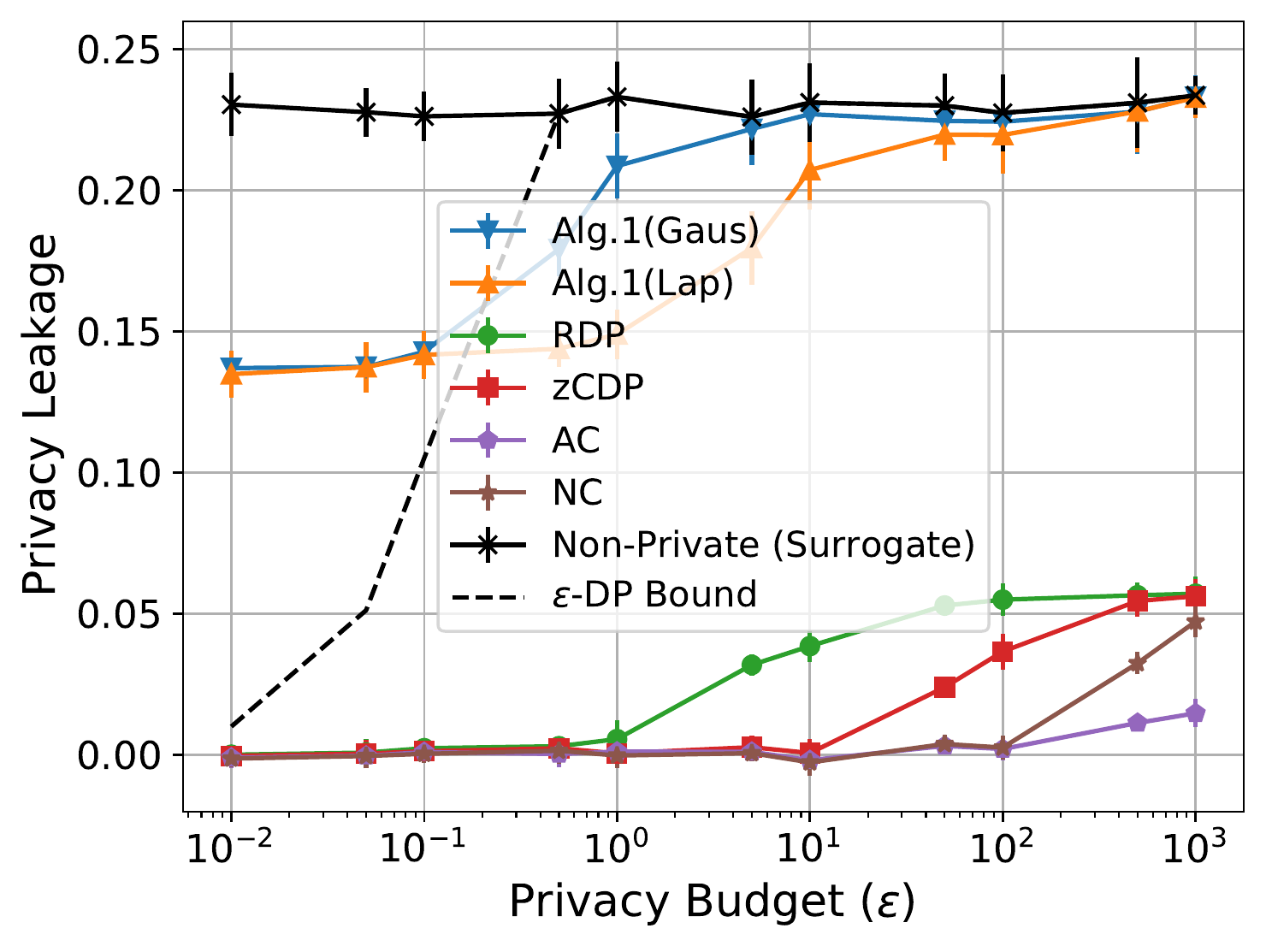}
    \label{subfig:texas_priv_leak}
    }
    \caption{Performance Evaluations on Texas Hospital Dataset.}
    \label{fig:texas}
\end{figure}

\begin{figure}[!ht]
    \centering
    \captionsetup{justification=centering}
    \subfloat[Accuracy Loss]{
    \includegraphics[width=0.23\textwidth]{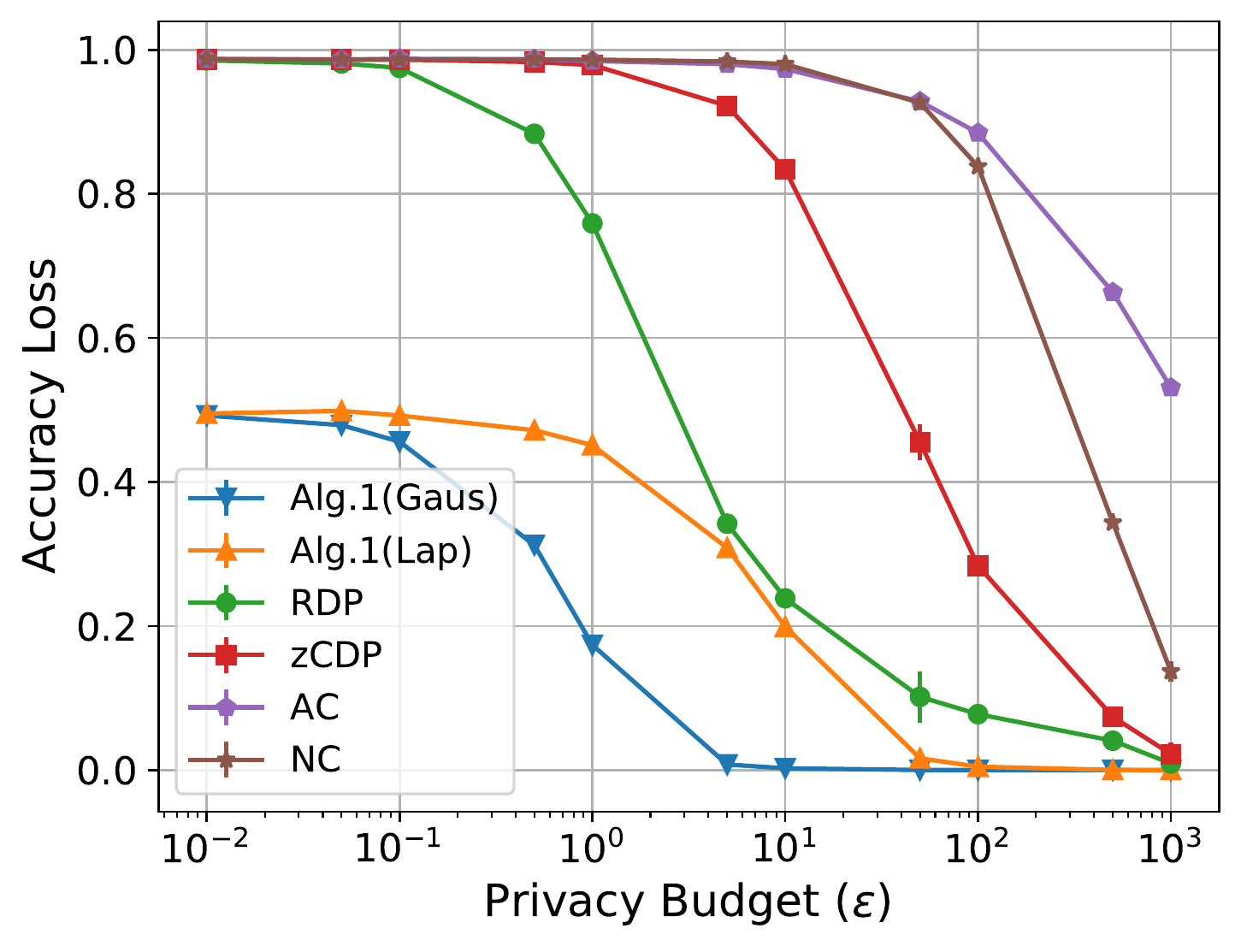}
    \label{subfig:purchase_100_acc_loss}
    }
    \hfill
    \subfloat[Privacy Leakage]{
    \includegraphics[width=0.23\textwidth]{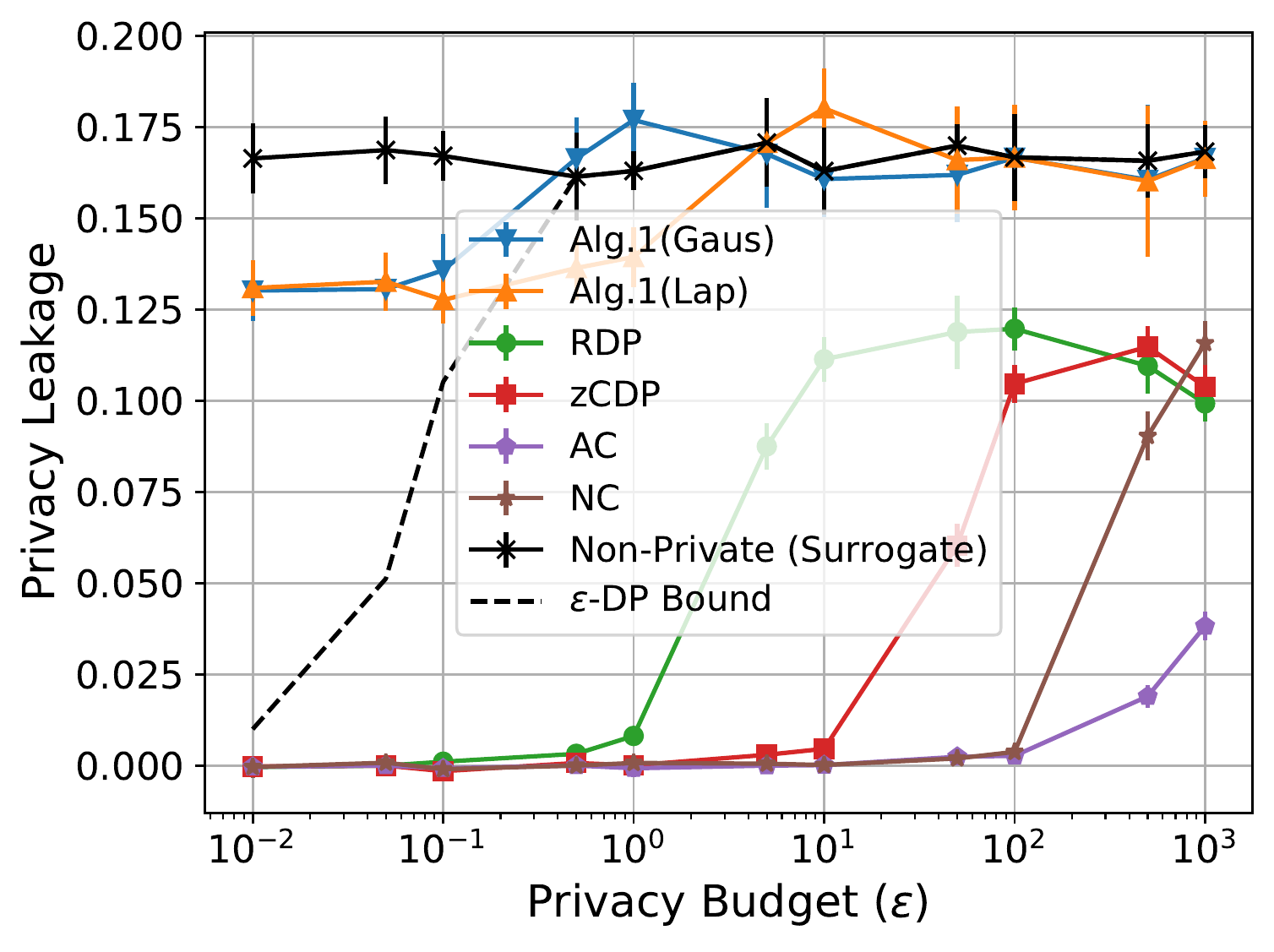}
    \label{subfig:purchase_100_priv_leak}
    }
    \caption{Performance Evaluations on Purchase-100 Dataset.}
    \label{fig:purchase_100}
\end{figure}

\begin{figure}[!ht]
    \centering
    \captionsetup{justification=centering}
    \subfloat[Accuracy Loss]{
    \includegraphics[width=0.23\textwidth]{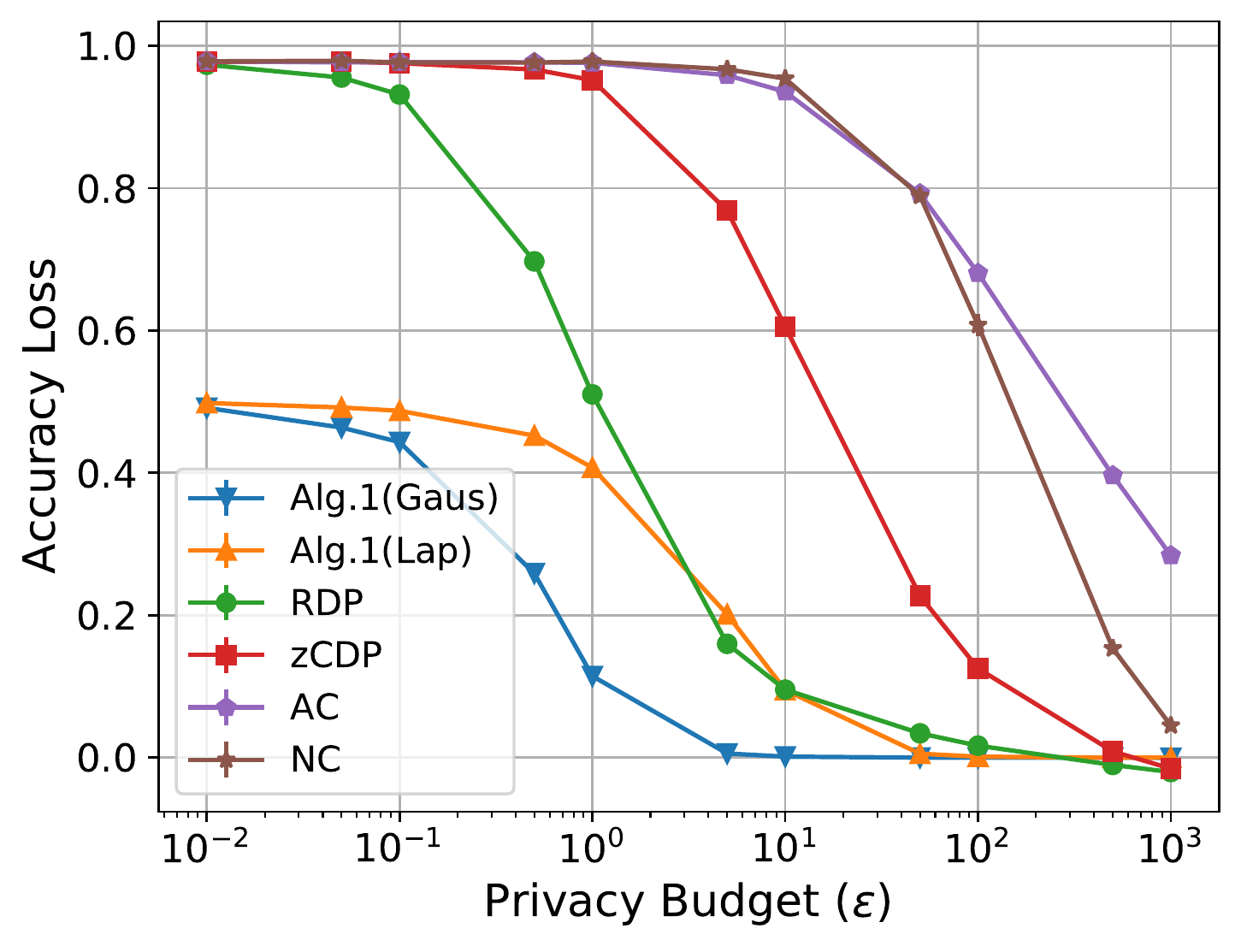}
    \label{subfig:purchase_50_acc_loss}
    }
    \hfill
    \subfloat[Privacy Leakage]{
    \includegraphics[width=0.23\textwidth]{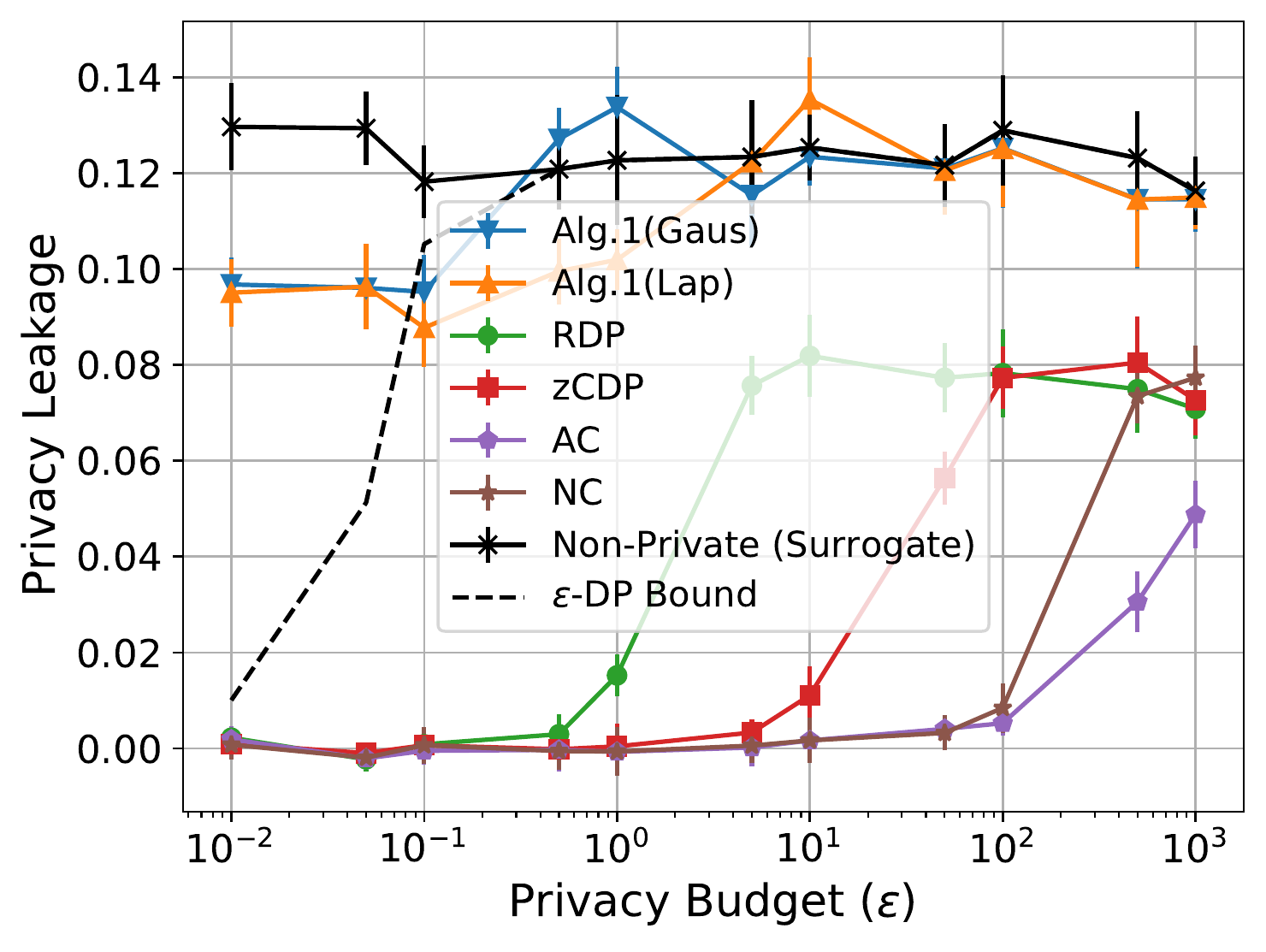}
    \label{subfig:purchase_50_priv_leak}
    }
    \caption{Performance Evaluations on Purchase-50 Dataset.}
    \label{fig:purchase_50}
\end{figure}

\begin{figure}[!ht]
    \centering
    \captionsetup{justification=centering}
    \subfloat[Accuracy Loss]{
    \includegraphics[width=0.23\textwidth]{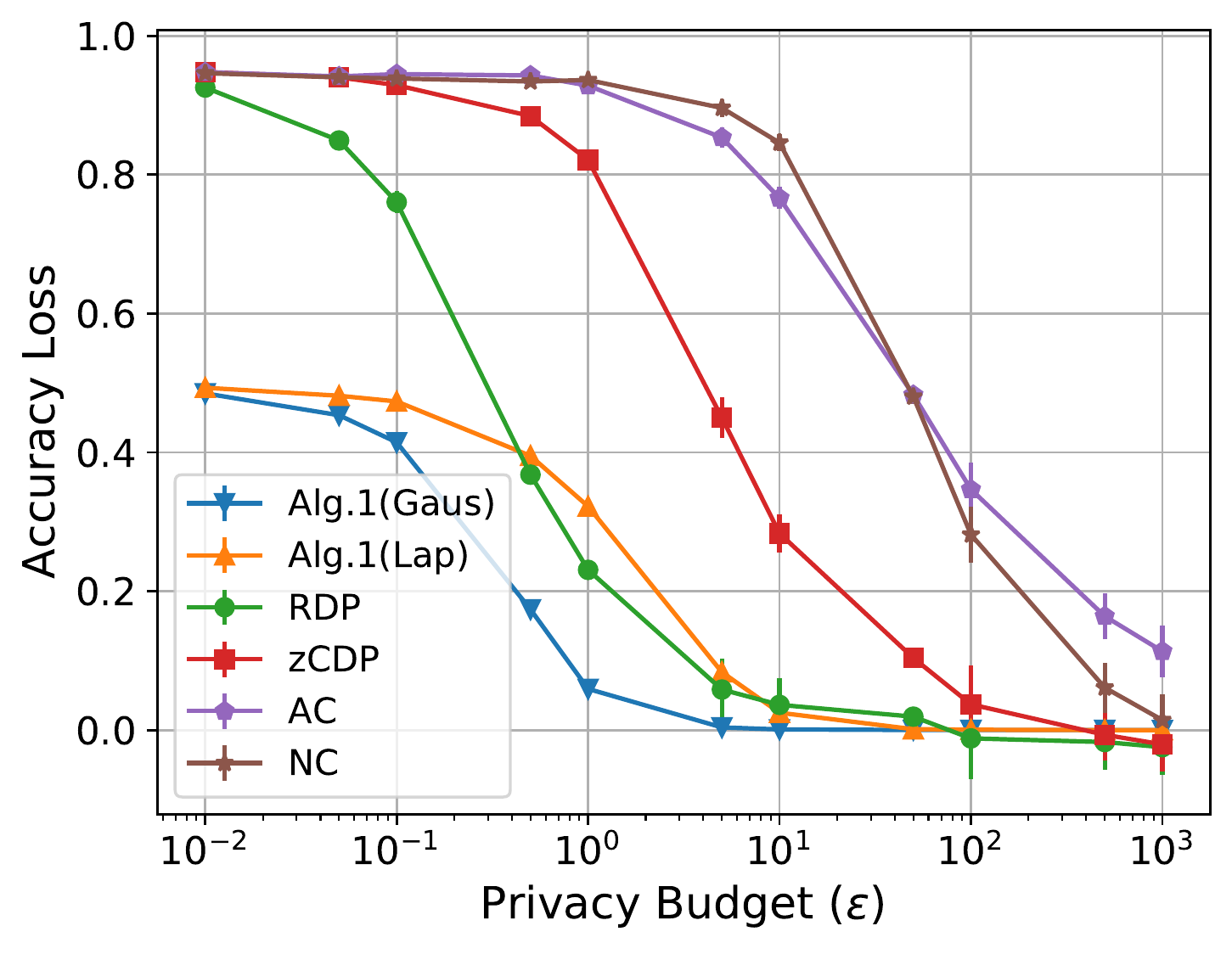}
    \label{subfig:purchase_20_acc_loss}
    }
    \hfill
    \subfloat[Privacy Leakage]{
    \includegraphics[width=0.23\textwidth]{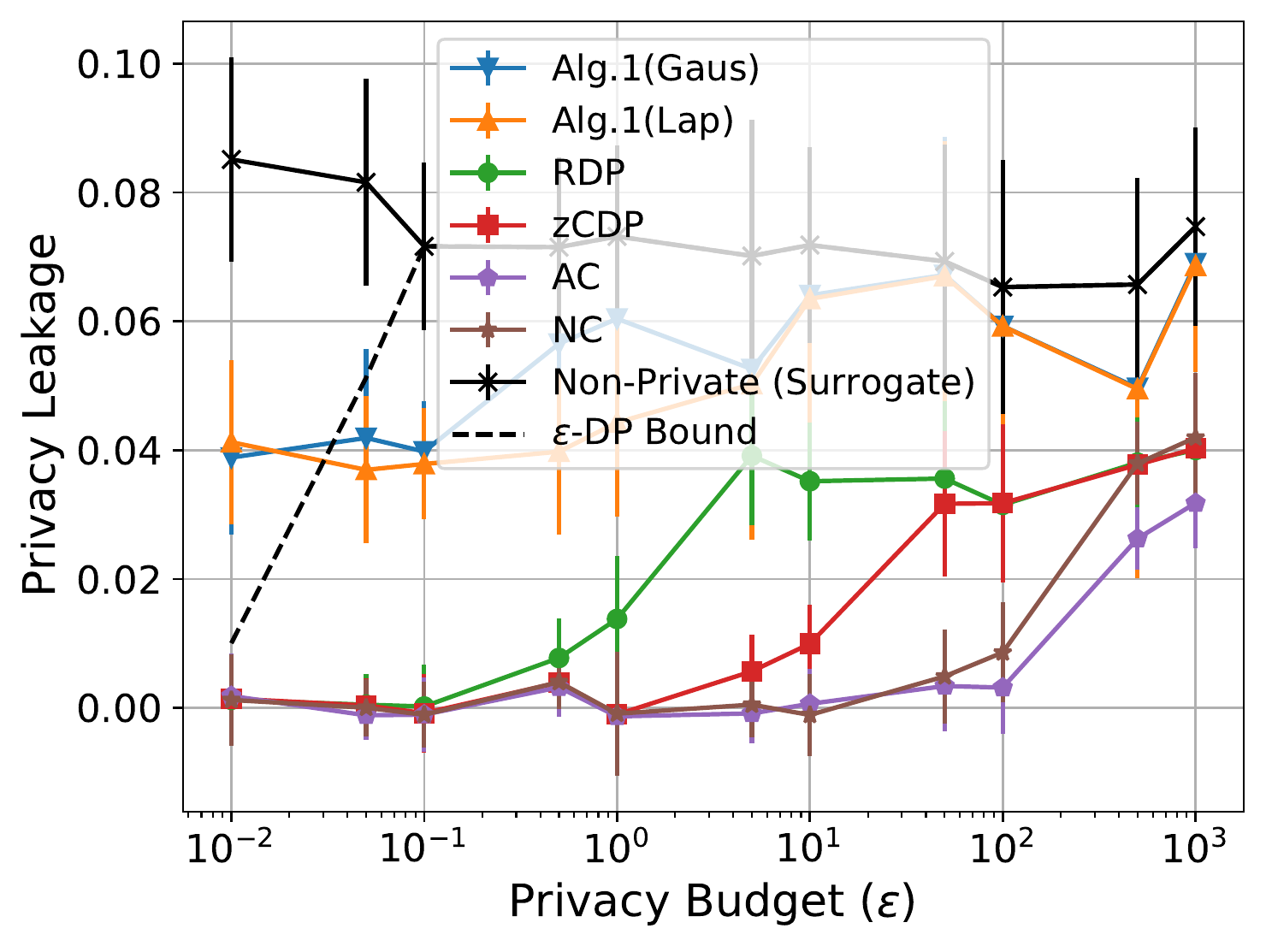}
    \label{subfig:purchase_20_priv_leak}
    }
    \caption{Performance Evaluations on Purchase-20 Dataset.}
    \label{fig:purchase_20}
\end{figure}

\begin{figure}[!ht]
    \centering
    \captionsetup{justification=centering}
    \subfloat[Accuracy Loss]{
    \includegraphics[width=0.23\textwidth]{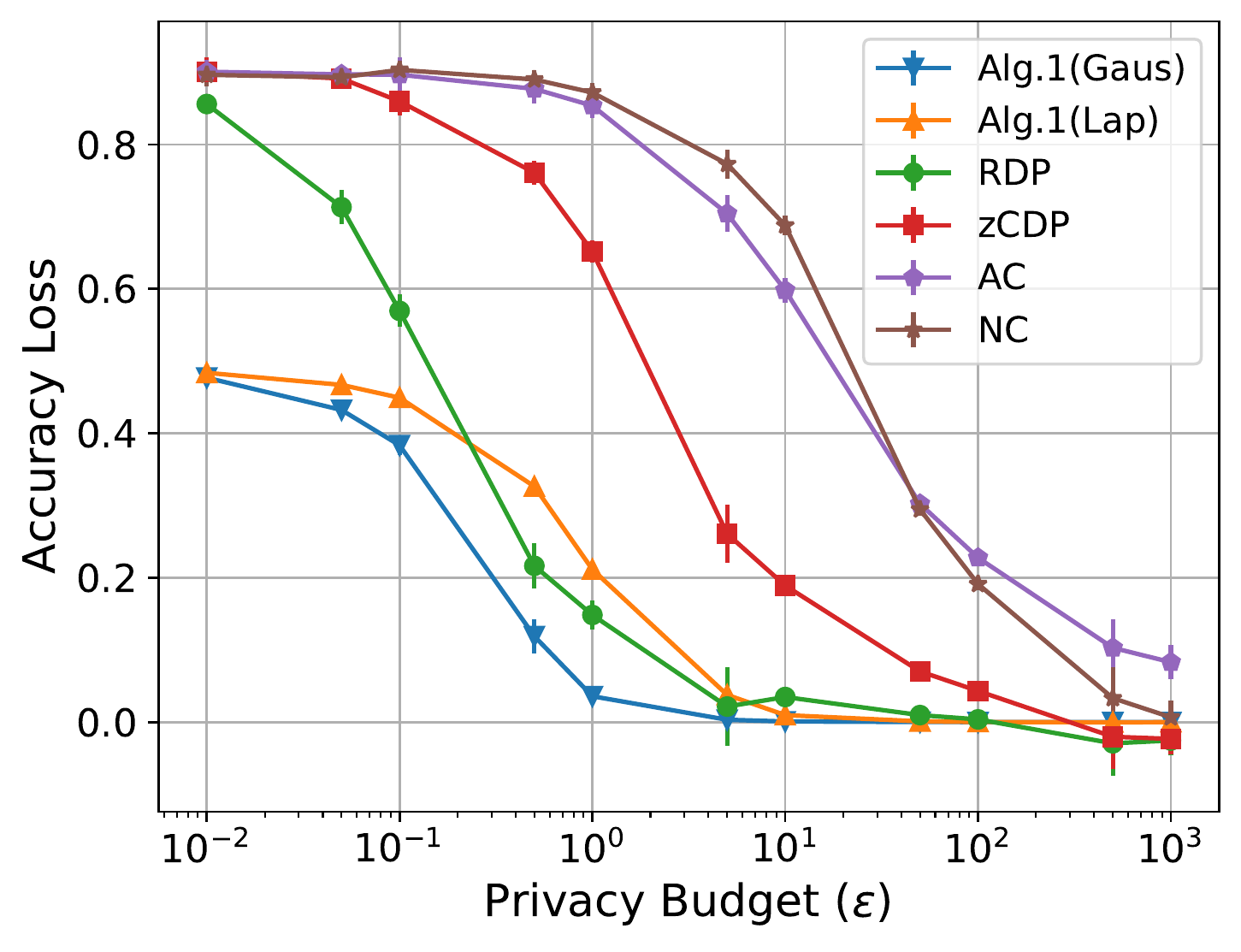}
    \label{subfig:purchase_10_acc_loss}
    }
    \hfill
    \subfloat[Privacy Leakage]{
    \includegraphics[width=0.23\textwidth]{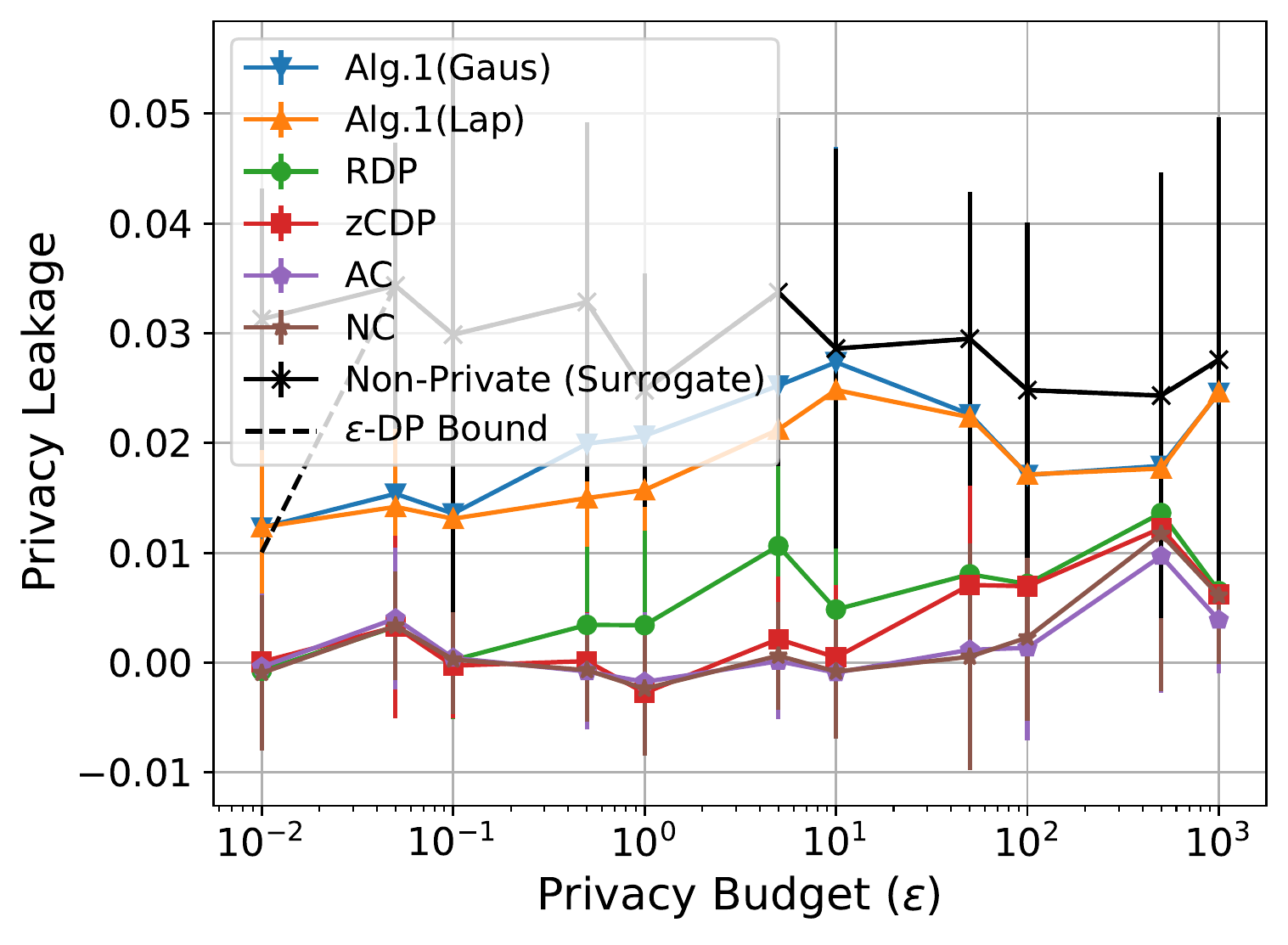}
    \label{subfig:purchase_10_priv_leak}
    }
    \caption{Performance Evaluations on Purchase-10 Dataset.}
    \label{fig:purchase_10}
\end{figure}

\begin{figure}[!ht]
    \centering
    \captionsetup{justification=centering}
    \subfloat[Accuracy Loss]{
    \includegraphics[width=0.23\textwidth]{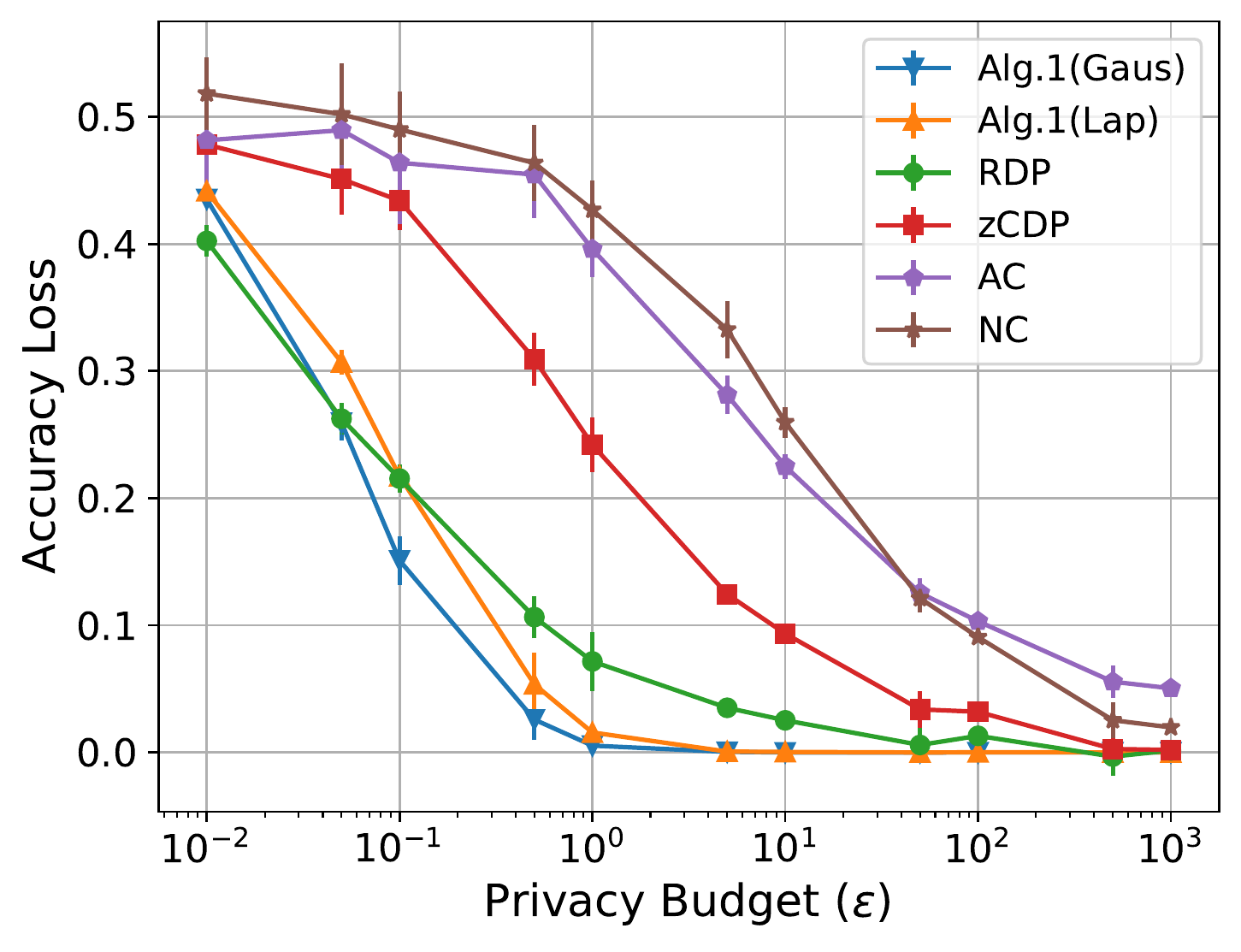}
    \label{subfig:purchase_2_acc_loss}
    }
    \hfill
    \subfloat[Privacy Leakage]{
    \includegraphics[width=0.23\textwidth]{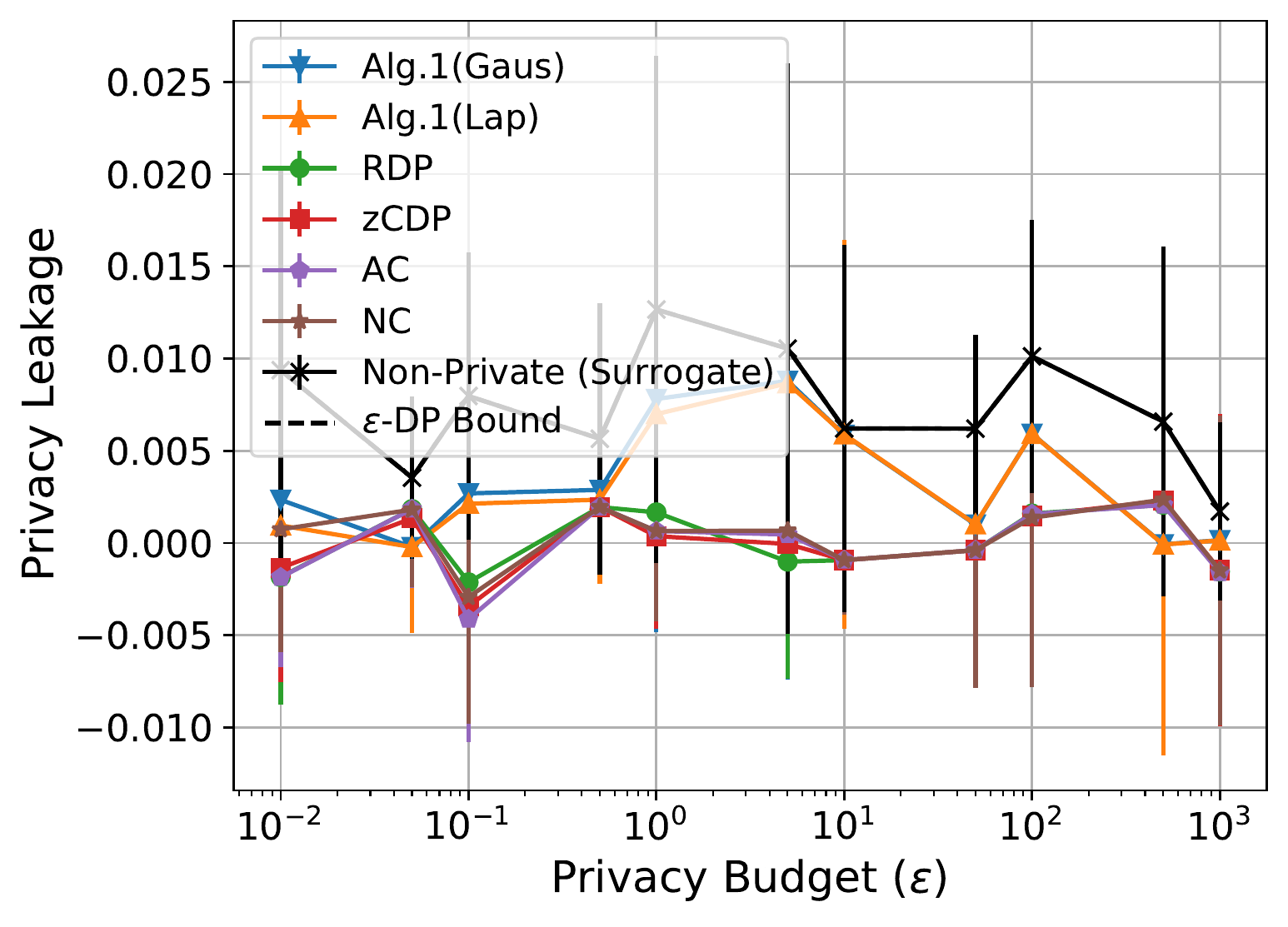}
    \label{subfig:purchase_2_priv_leak}
    }
    \caption{Performance Evaluations on Purchase-2 Dataset.}
    \label{fig:purchase_2}
\end{figure}

\begin{figure}[!ht]
    \centering
    \captionsetup{justification=centering}
    \subfloat[Accuracy Loss]{
    \includegraphics[width=0.23\textwidth]{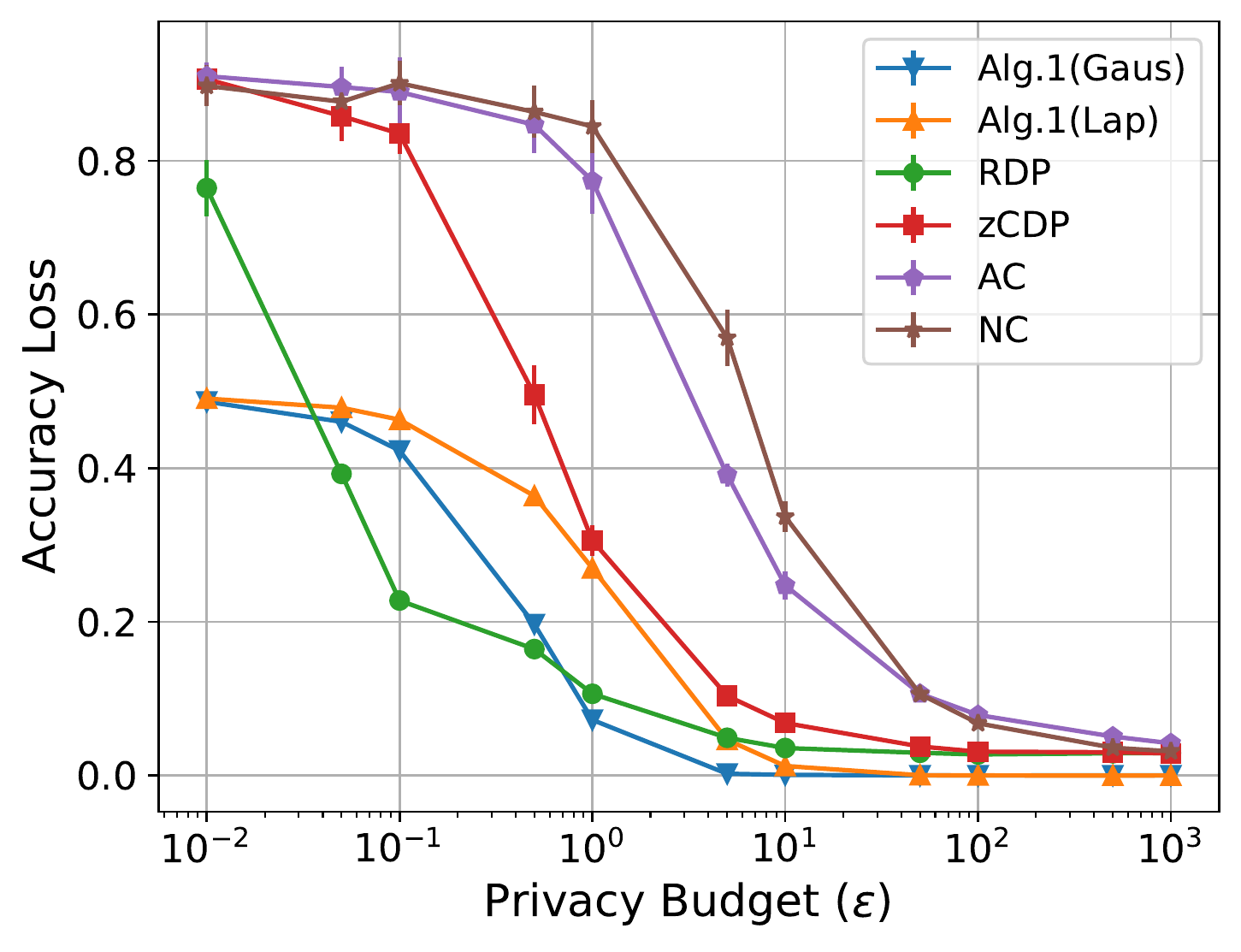}
    \label{subfig:mnist_10_acc_loss}
    }
    \hfill
    \subfloat[Privacy Leakage]{
    \includegraphics[width=0.23\textwidth]{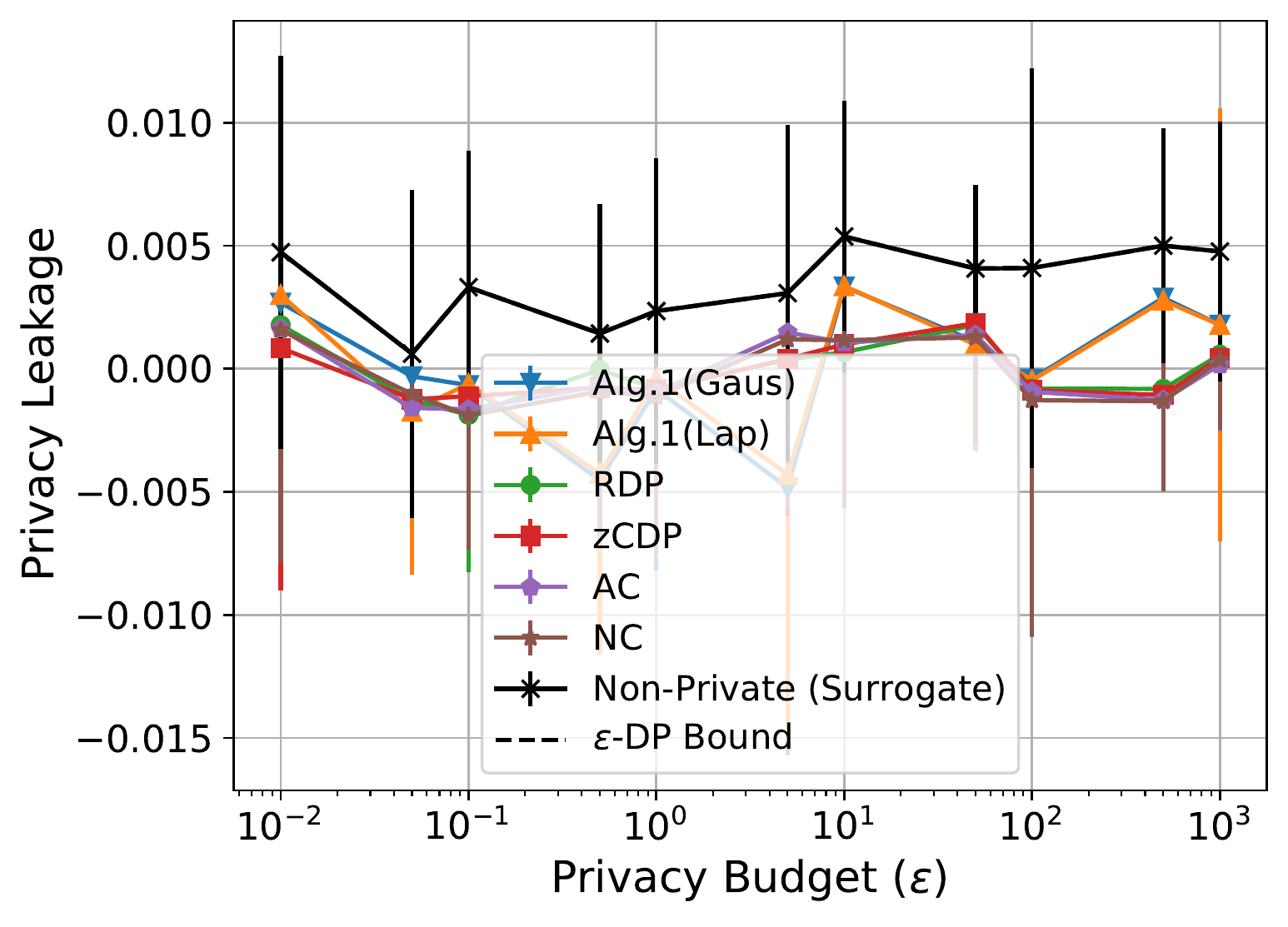}
    \label{subfig:mnist_10_priv_leak}
    }
    \caption{Performance Evaluations on MNIST Dataset.}
    \label{fig:mnist}
\end{figure}

\begin{figure}[!ht]
    \centering
    \captionsetup{justification=centering}
    \subfloat[Accuracy Loss]{
    \includegraphics[width=0.23\textwidth]{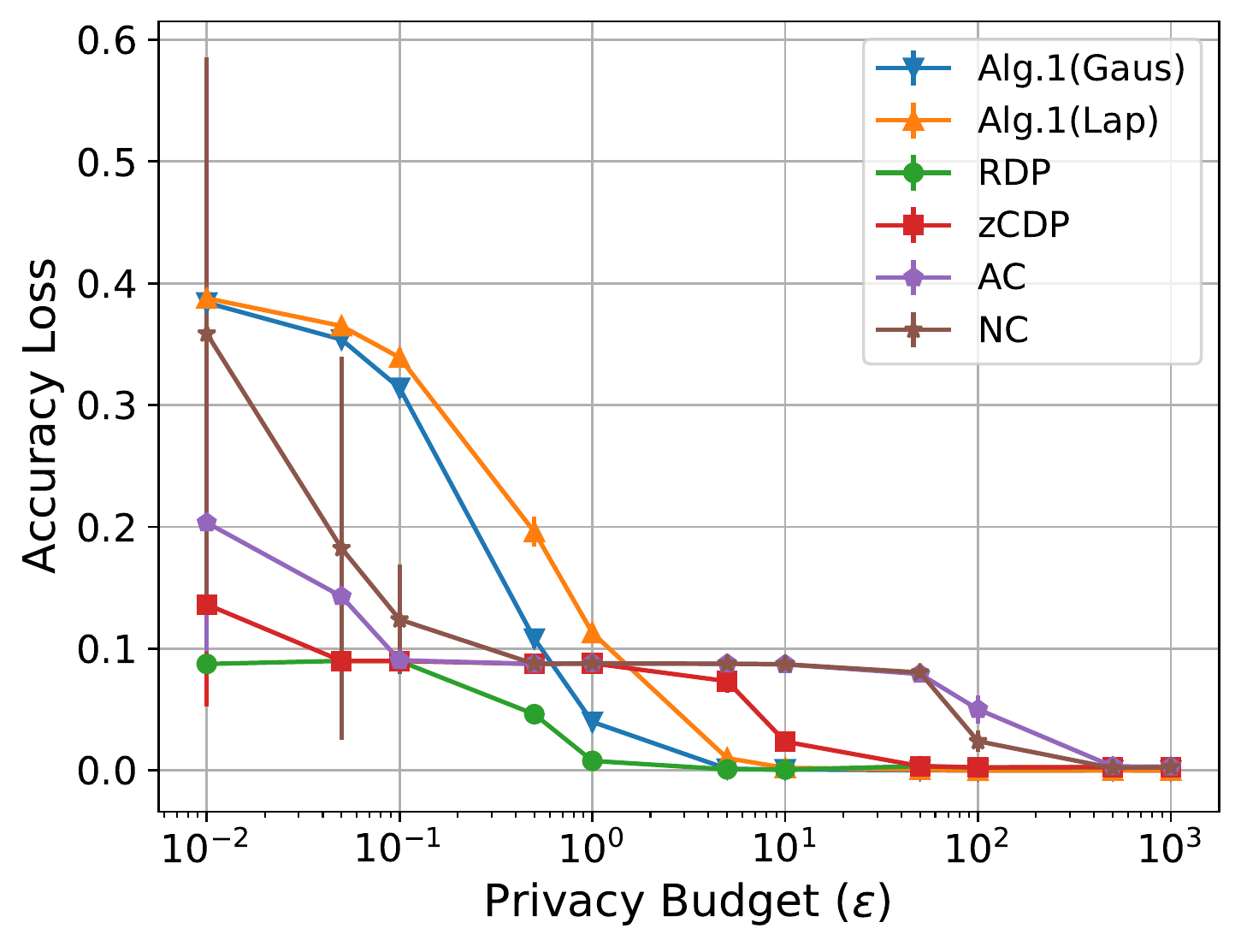}
    \label{subfig:adult_acc_loss}
    }
    \hfill
    \subfloat[Privacy Leakage]{
    \includegraphics[width=0.23\textwidth]{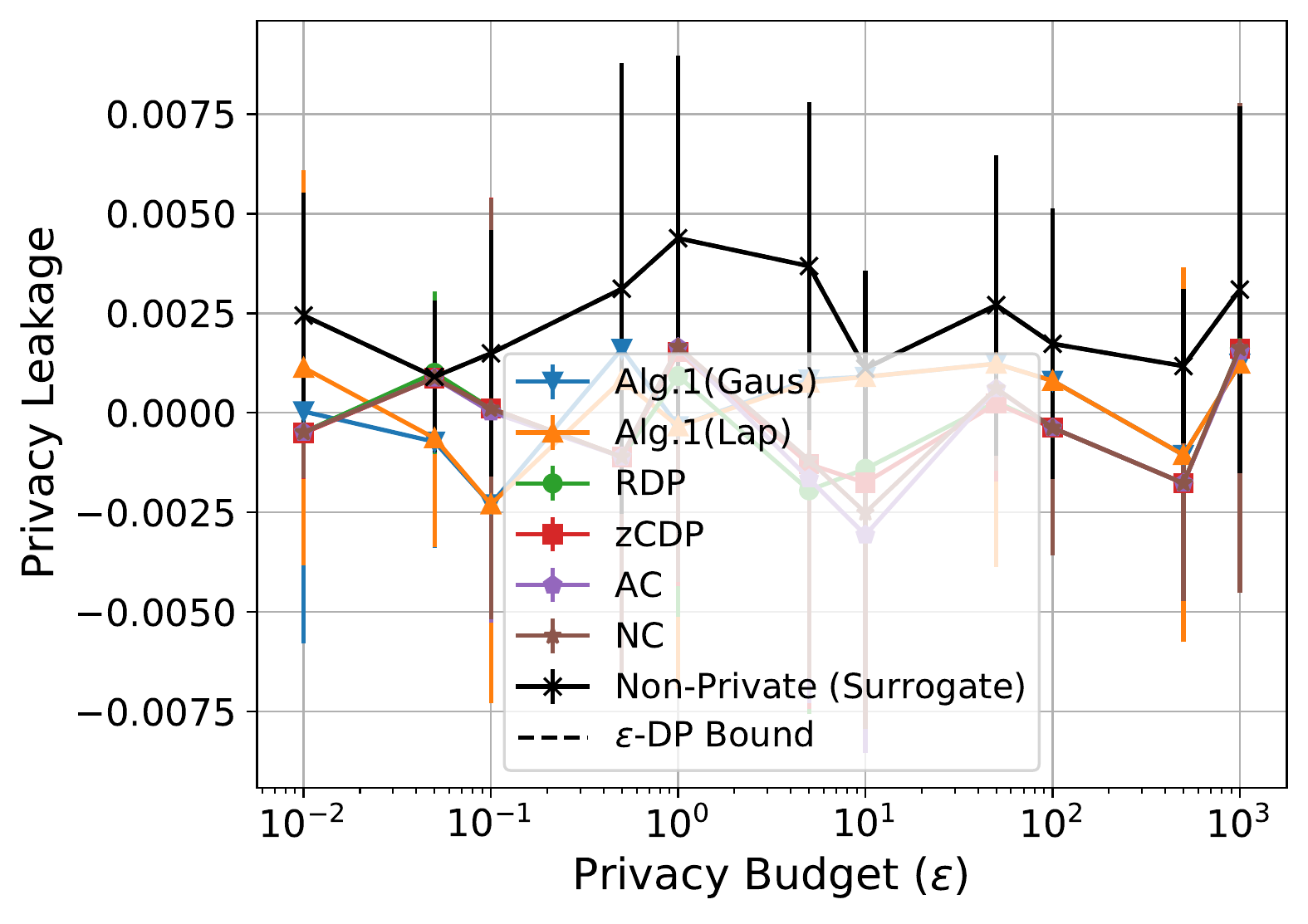}
    \label{subfig:adult_priv_leak}
    }
    \caption{Performance Evaluations on US Adult Dataset.}
    \label{fig:adult}
\end{figure}

\begin{figure}[!ht]
    \centering
    \captionsetup{justification=centering}
    \subfloat[Accuracy Loss]{
    \includegraphics[width=0.23\textwidth]{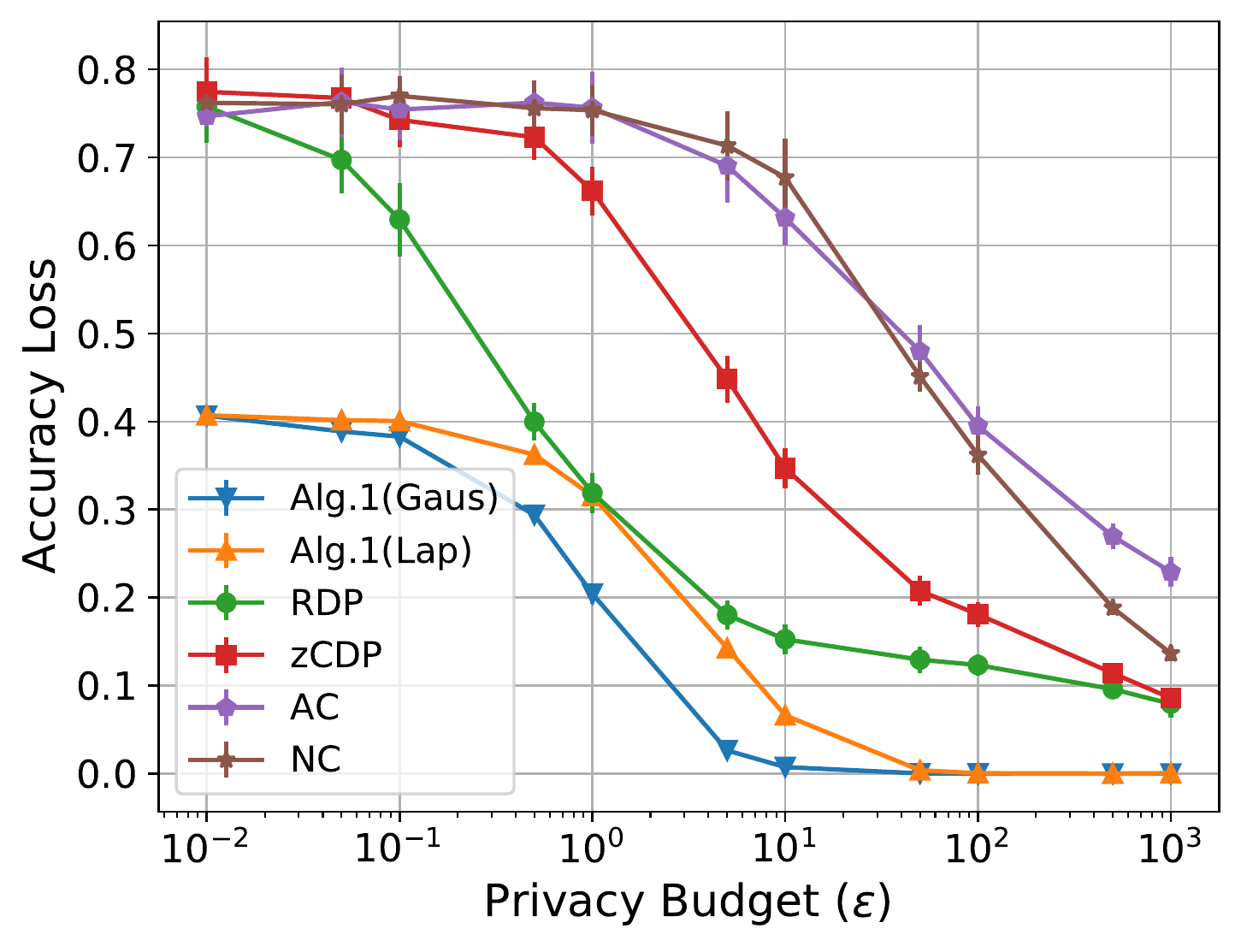}
    \label{subfig:cifar_10_acc_loss}
    }
    \hfill
    \subfloat[Privacy Leakage]{
    \includegraphics[width=0.23\textwidth]{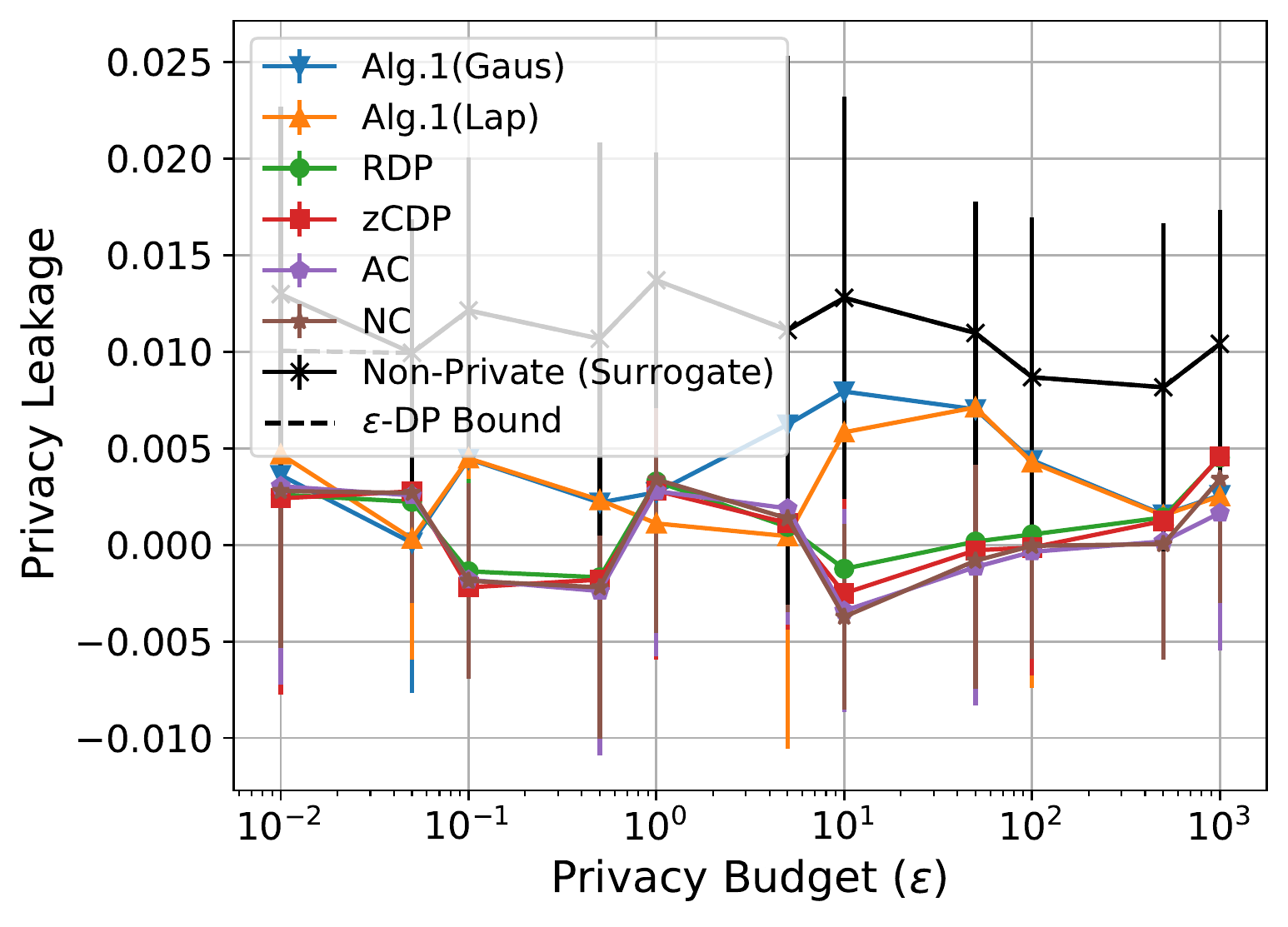}
    \label{subfig:cifar_10_priv_leak}
    }
    \caption{Performance Evaluations on CIFAR-10 Dataset.}
    \label{fig:cifar_10}
\end{figure}

\begin{figure}[!ht]
    \centering
    \captionsetup{justification=centering}
    \subfloat[Accuracy Loss]{
    \includegraphics[width=0.23\textwidth]{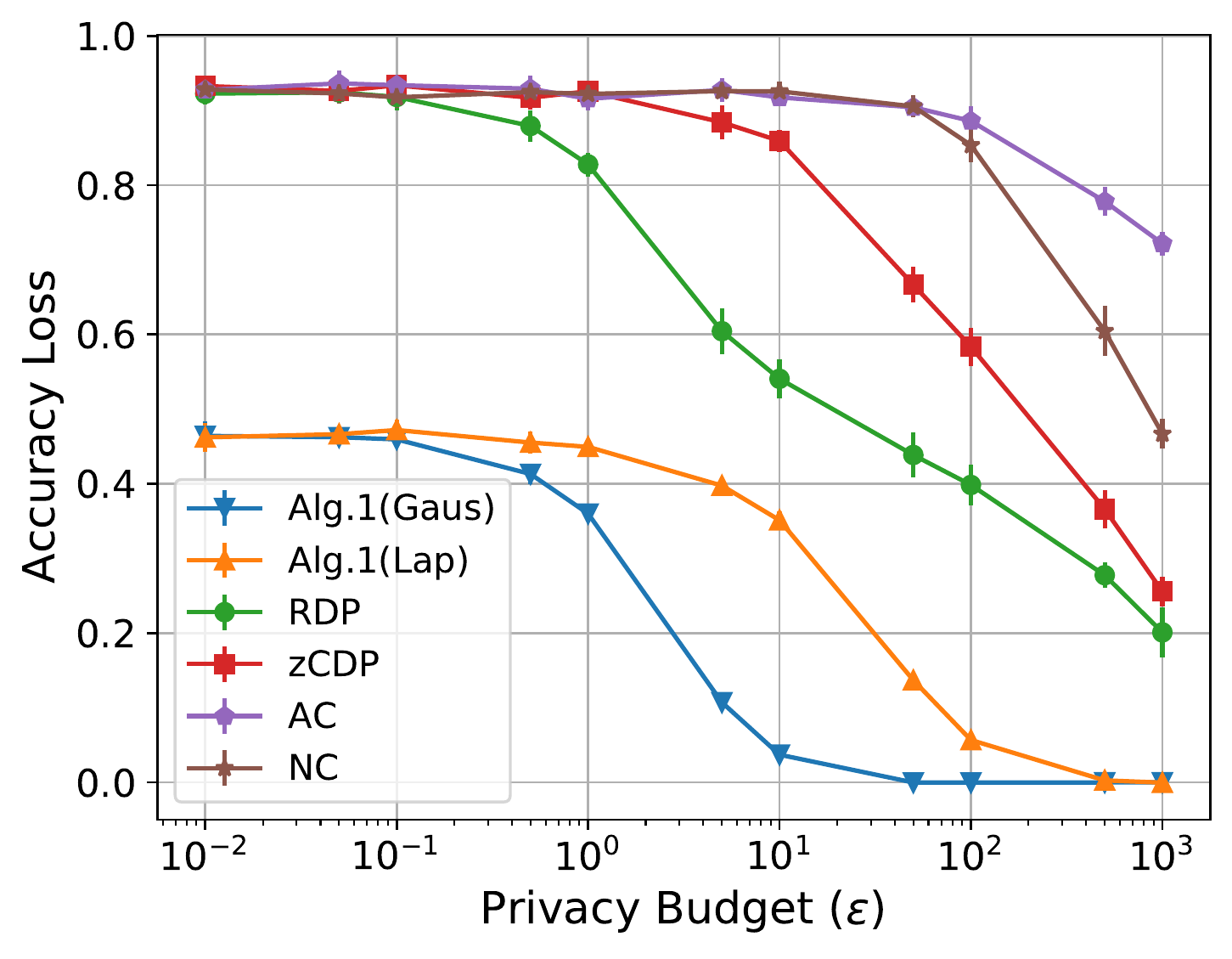}
    \label{subfig:cifar_100_acc_loss}
    }
    \hfill
    \subfloat[Privacy Leakage]{
    \includegraphics[width=0.23\textwidth]{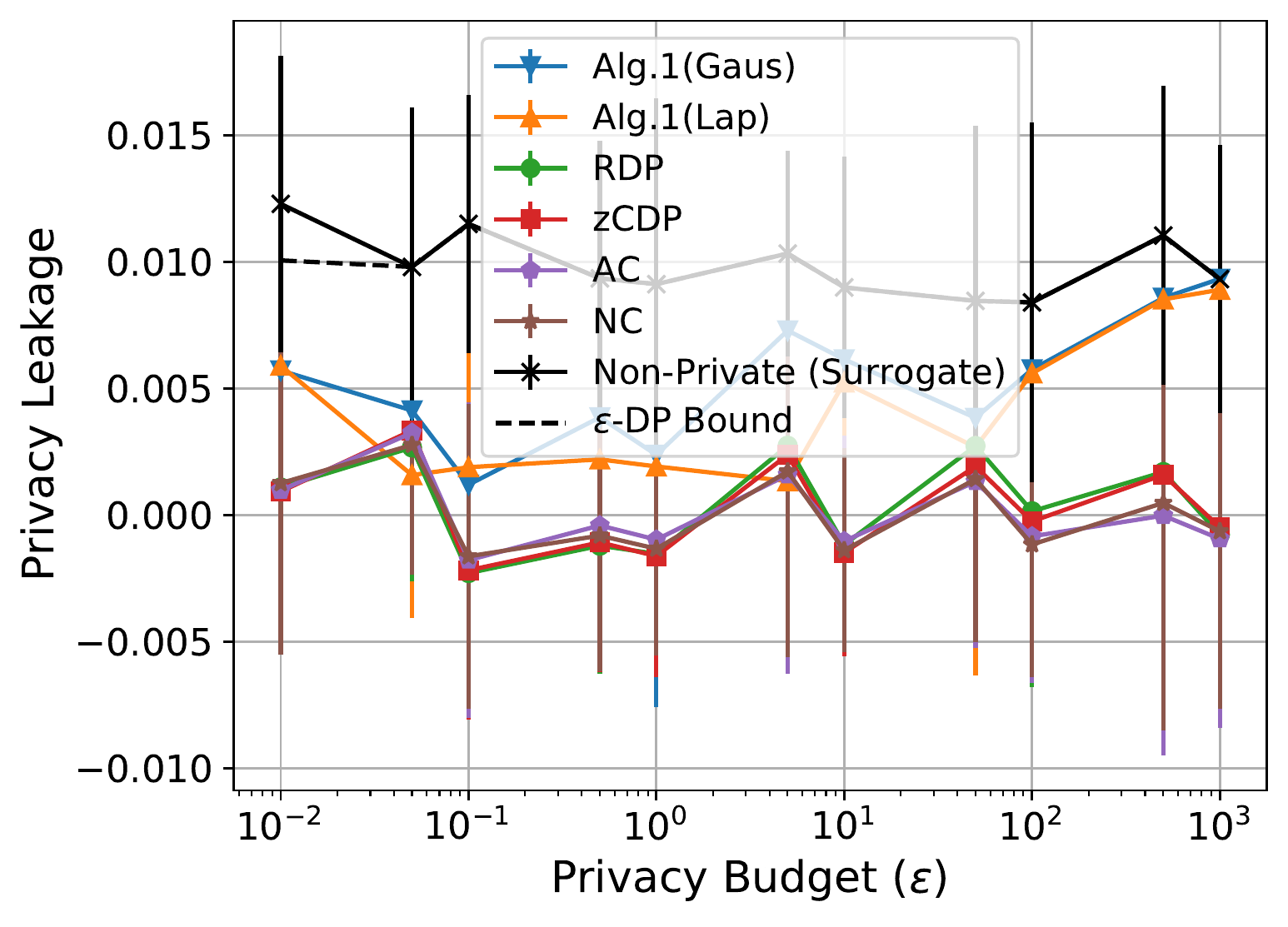}
    \label{subfig:cifar_100_priv_leak}
    }
    \caption{Performance Evaluations on CIFAR-100 Dataset.}
    \label{fig:cifar_100}
\end{figure}

Figures~\ref{fig:location} to \ref{fig:cifar_100} depict the accuracy loss and privacy leakage on the six real-world datasets, where \textit{Alg.~\ref{alg:dpdnn} (Gaus)} and \textit{Alg.~\ref{alg:dpdnn} (Lap)} represent Algorithm~\ref{alg:dpdnn} implemented with Gaussian noise and Laplace noise, respectively. \reviseA{Table~\ref{tab:sensitivity} shows the Lipschitz constant and the global sensitivities we calculated and used in our experiments based on our theoretical results.} We give our key findings below.

\textbf{Finding 1: Avoiding overfitting is still the rule of thumb to mitigate the effect of MI attacks (via shadow models).} Based on the observed accuracy gap and privacy leakage of the non-private models on different datasets (Figure~\ref{subfig:priv_leak_acc_gap}, where the Pearson correlation coefficient is 0.9592, Spearman's rank and Kendall's rank correlation coefficients are 1, the p-value of all the three correlation coefficient is 0), we have the same conclusion as existing works~\cite{YeomS2018,TruexS2019,TonniS2020}. That is, when a model is not overfitting, the privacy leakage of the non-private model would be rather marginal (almost zero privacy leakage) against MI attacks. 


\textbf{Finding 2: The accuracy loss of Algorithm~\ref{alg:dpdnn} follows a stable monotonic decreasing curve when tuning the privacy budget.} This property gives us a predictable accuracy loss when configuring a specific value of the privacy budget to generate DP prediction. Such a stability is in two-fold. First, on a single query, the maximum accuracy loss of Algorithm~\ref{alg:dpdnn} is about $0.5$, which is much smaller than existing DP-SGD approaches. Second, the accuracy loss of Algorithm~\ref{alg:dpdnn} (Gaussian noise) significantly decreases from its maximum value to $0$ when $\epsilon \in [0.1, 10]$, which is the commonly used range for tuning the privacy budget in practice (for Laplace noise, $\epsilon \in [1, 100]$). We observe exceptions on Location (Figure~\ref{fig:location}) and Purchase-2 (Figure~\ref{fig:purchase_2}) datasets, where such ranges of the privacy budget are in $[1, 100]$ and $[0.01, 1]$, respectively. Whereas, we cannot observe such a stable curve in existing DP-SGD approaches in any aspect, such as the maximum accuracy loss (varying from $1.0$ to $0.1$ on different datasets), range of privacy budget for decreasing the accuracy loss (varying from $[0.05, 100]$ to $[0.01, 1]$ on different datasets).

\textbf{Finding 3: Algorithm~\ref{alg:dpdnn} achieves a good privacy-utility trade-off when the privacy leakage of the baseline model does not approximately equal to 0.} Specifically, in most datasets (except the two most fitted models on MNIST and US Adult datasets, Figures~\ref{fig:mnist} and \ref{fig:adult}, where the privacy leakage is less than $0.02$), for a given privacy budget, Algorithm~\ref{alg:dpdnn} (Gaussian noise and Laplace noise) provides the least accuracy loss and achieves the closest privacy leakage to the theoretical bound on $\epsilon$-DP for a single query. When tuning the privacy budget to large values, say over $10$ (Gaussian noise) or $100$ (Laplace noise), Algorithm~\ref{alg:dpdnn} could converge to the same privacy leakage as the non-private models in most of the datasets, which is expected as a DP algorithm.


\textbf{Finding 4: Algorithm~\ref{alg:dpdnn} (Laplace noise) achieves similar accuracy loss curve as Algorithm~\ref{alg:dpdnn} (Gaussian noise) for the binary classes datasets.} In the two binary classes datasets, Purchase-2 (Figure~\ref{fig:purchase_2}) and US Adult (Figure~\ref{fig:adult}), the performance of injecting Gaussian noise and injecting Laplace noise is close to each other. Whereas, on the multi-class datasets, injecting Gaussian noise always outperforms Laplace noise injection on the privacy-utility trade-off.

\reviseA{
\textbf{Finding 5: For a given neural network topology, the upper bounds on the global sensitivities are highly depending to the number of classes and the number of records of the training set, rather than the Lipschitz constant.} From the results reported in Table~\ref{tab:sensitivity}, the values of Lipschitz constant $\rho$ varies over different datasets; whereas the global sensitivities $\Delta \omega$ and $\Delta z_{T}$ remain relatively stable. Exceptions on the global sensitivities are US Adult, Purchase-2 and Location datasets, where the former two are binary classes datasets (where the global sensitivities are about half of the global sensitivities of the datasets having the same training set size) and the latter contains 1200 records (whose global sensitivities are about 10 times more than the global sensitivities of datasets having multiple classes).
}

\begin{table}[!th]
\centering
\caption{\reviseA{Sensitivities and Lipschitz Constant of Alg.~\ref{alg:dpdnn} Calculated and Used in Experiments.}}
\label{tab:sensitivity}
\scalebox{0.85}{
\begin{tabular}{l|c|c|c|c}
\hline
Dataset & $\rho$ & $\Delta \omega$ & $\Delta z_{T}$ & $\Delta p$ ($\exp(2\Delta z_{T})-1$) \\ \hline\hline
US Adult & 0.1654 & 0.0015 & 0.1871 & 0.4538 (0.4538) \\ \hline
MNIST & 2.2274 & 0.0028 & 0.3577 & 1.0000 (1.0450) \\ \hline
Location & 1.8044 & 0.0244 & 3.1190 & 1.0000 (510.8338) \\ \hline
Purchase-2 & 1.0771 & 0.0016 & 0.2000 & 0.4918 (0.4918) \\ \hline
Purchase-10 & 1.9388 & 0.0028 & 0.3570 & 1.0000 (1.0421) \\ \hline
Purchase-20 & 2.0465 & 0.0029 & 0.3738 & 1.0000 (1.1119) \\ \hline
Purchase-50 & 2.1111 & 0.0029 & 0.3765 & 1.0000 (1.1234) \\ \hline
Purchase-100 & 2.1327 & 0.0029 & 0.3664 & 1.0000 (1.0809) \\ \hline
CIFAR-10 & 4.4091 & 0.0028 & 0.3594 & 1.0000 (1.0520) \\ \hline
CIFAR-100 & 4.8500 & 0.0030 & 0.3897 & 1.0000 (1.1802) \\ \hline
Texas Hospital & 6.8729 & 0.0031 & 0.3928 & 1.0000 (1.1937) \\ \hline
\end{tabular}
}
\end{table}

\subsubsection{Analysis of Experimental Results}
The observations in Section~\ref{sec:exp_surrogate} and Section~\ref{sec:exp_stability}, reflect the theoretical properties of the \revise{convexified} loss function shown in \cite{LoJ2012,DvijothamK2014} and the OARO stability in Definition~\ref{def:remove_stable} and Corollary~\ref{cr:dp_effect}. 

Findings 1, 2, 3, 4 \reviseA{and 5} in Section~\ref{sec:exp_dp} are supported by the noise injection scheme of Algorithm~\ref{alg:dpdnn}, i.e., applying Exp-DP to sample an individual neuron for noise injection with a tight upper bound on the global sensitivity could provide a better privacy-utility trade-off for an output perturbation-based DP algorithm. 


First, based on the measurement for model's accuracy, injecting large positive noise into the neuron representing the accurate prediction class (or injecting small negative noise into the neurons representing the inaccurate prediction classes) would not impact the test accuracy of Algorithm~\ref{alg:dpdnn}. When $\epsilon \leq 10^{-2}$, the sampling probabilities (applying Exp-DP) of all neurons at the output layer are similar. In this case, Algorithm~\ref{alg:dpdnn}, injecting noise following Laplace distribution or Gaussian distribution, gives the same (or similar) prediction as the baseline model half the time. Hence, we observe $\text{Acc\_Loss} \approx 0.5$ and $\text{Priv\_Leak} > 0$ from Algorithm~\ref{alg:dpdnn} for $\epsilon \leq 10^{-2}$. Once having a relatively greater privacy budget, the neuron representing the accurate prediction class would have much higher probability to be sampled, together with the tight upper bound on the global sensitivities, Algorithm~\ref{alg:dpdnn} ensures a better privacy-utility trade-off for a single query.

Second, in the binary-class datasets, since there are only two neurons at the output layer, with high probability, the Exp-DP would sample the neuron representing the accurate prediction class. In this case, the amount of noise would not impact the final prediction outcome. Hence, we cannot observe significant differences between Gaussian noise and Laplace noise in binary-class datasets. 

\reviseA{
Third, as discussed in Section~\ref{sec:lip_cons}, roughly $\frac{C-1}{Cn}$ is the factor impacting the global sensitivities, where $C$ is the number of classes and $n$ is the number of training records. $\frac{C-1}{Cn}$ is a monotonically increasing function for $C \in [2, +\infty)$ and a monotonically decreasing function for $n \in [1, +\infty)$. When $C = 2$ (binary datasets) we get $\frac{C-1}{C} = \frac{1}{2}$. On the other hand, when $C \geq 10$, we get $\frac{C-1}{C} \approx 1$. This implies that the global sensitivities obtained in US Adult and Purchase-2 datasets (binary class datasets) are about half of that in other datasets ($C > 2$, $n = 10,000$) as can be seen in Table~\ref{tab:sensitivity}. On the other hand, the Location dataset is the only dataset having $n = 1,200$ records, as compared to $n = 10,000$ records for all other datasets (see Table~\ref{tab:datasets}). As a result, the global sensitivities obtained in Location dataset are about 10 times than that in other datasets ($C > 2$, $n = 10,000$).
}

\reviseA{
\subsubsection{Privacy Budget Consumption for Multiple Queries.}
\label{sub:budget-multiple-queries}
Since Algorithm~\ref{alg:dpdnn} is based on output perturbation, it consumes privacy budget for each query (a single data sample for prediction), and hence can only be used for a fixed number of queries before exhausting the privacy budget. On the other hand, DP-SGD, as a gradient perturbation, consumes privacy budget during the training process, so it could accept an unlimited number of queries without further privacy budget consumption. However, we can always scale the privacy budget according to the number of queries we are willing to answer.

Based on the experimental results observed in Figures~\ref{fig:location} to \ref{fig:purchase_50}, we conclude that when the privacy leakage of a non-private model is no less than $0.1$, given an overall privacy budget $\epsilon \leq 1$ for multiple queries, Algorithm~\ref{alg:dpdnn} (Gaussian) could answer a large number of queries (which is likely to be the size we expect in practice), while still outperforming DP-SGD in the privacy-utility trade-off. Note that while we report these results for single queries, we can use the same results assuming a larger number of queries for a fixed budget. For instance, consider the results on the Location dataset in Figure~\ref{fig:location}. If we have $\epsilon = 10$, and we would like to answer 1,000 queries, then we can look at the accuracy loss (privacy leakage) at $\epsilon = 10^{-2} = 10/1000$ in the figure, for the accuracy loss (privacy leakage) per query.  

Thus, again on the Location dataset (Figure~\ref{fig:location}), given $\epsilon = 10$ for $100$ queries, based on the sequential composition of privacy budget (Theorem~\ref{thm:Lap-DP_composition}), each single query consumes $\epsilon = 10^{-2}$. At $\epsilon = 10^{-2}$, Algorithm~\ref{alg:dpdnn} (Gaussian) has about $0.5$ accuracy loss and $0.2$ privacy leakage, whereas the best DP-SGD (RDP) at $\epsilon = 1$ (since DP-SGD does not consume privacy budget during the test/prediction phase) has about $0.9$ accuracy loss and $0.0$ privacy leakage. That is, Algorithm~\ref{alg:dpdnn} has less accuracy loss and more privacy leakage, hence having a better privacy-utility trade-off in this example. For a larger number of queries (1,000 or 10,000) the privacy-utility loss is comparable to DP-SGD. Based on the analysis of Finding 2, for a single query, when $\epsilon < 10^{-2}$, Algorithm~\ref{alg:dpdnn} should still have $\text{Acc\_Loss} \approx 0.5$ and $\text{Priv\_Leak} > 0$, then we could extend the number of queries answered by Algorithm~\ref{alg:dpdnn} to be a large number in practice. Thus, the advantage of our method is that we can provide higher accuracy if the model is required to answer a small number of queries, which is not the case with input perturbation or gradient perturbation-based methods such as DP-SGD.}

\section{Conclusion}
\label{sec:conclusion}
\revise{\textbf{Concluding Remarks.}} In this paper we propose a framework that provides differentially private prediction probability vector for general deep learning tasks. Our approach only injects DP noise into one neuron (sampled with Exponential mechanism of differential privacy) at the output layer of a given neural network. To implement our approach, we mathematically analyse the upper bound on $L_{1}$ global sensitivity of an individual neuron via analysing a tighter upper bound on $L_{2}$ global sensitivity of the trained model parameters (than existing results). Our empirical studies show that our approach achieves a better trade-off between utility and privacy than existing DP-SGD approaches on six commonly used real-world datasets, \reviseA{given an overall privacy budget $\epsilon \leq 1$ for a large number of queries.}

\revise{
\textbf{Limitations and Future Work.} Our approach only provides DP predictions before reaching a pre-defined number of queries, since we consume privacy budget per query. For an output perturbation-based solution, to answer an unlimited number of queries while guaranteeing DP, it requires injecting DP noise directly into all the trained model parameters. However, due to the complexity of the topology of neural networks, this would result in over-injected noise adversely impacting utility of prediction. Additionally, a tight upper bound on $L_{2}$ global sensitivity of the trained model parameters does not always give us a tight upper bound on $L_{1}$ global sensitivity of an individual neuron as shown in Section~\ref{sec:global_sens}. Instead, experimental results show trivial upper bounds in most datasets (Table~\ref{tab:sensitivity}) having no greater than $10,000$ training data samples. To improve our results, i.e., answering more queries, we should consume less privacy budget per query by either exploring other noise injection schemes having better privacy budget composition or further tightening our upper bounds on the global sensitivity. This is an open question for future work.
}

\section*{Acknowledgements}
We thank Bargav Jayaraman for clarifications on the use of their implementation of DP-SGD. This work was conducted with funding received from the Macquarie University Cyber Security Hub, in partnership with the Defence Science \& Technology Group and Data61-CSIRO, through the Next Generation Technologies Fund. The experiments of this work was partially supported by the Australasian Leadership Computing Grants scheme, with computational resources provided by NCI Australia, an NCRIS enabled capability supported by the Australian Government. Hassan Jameel Asghar is the corresponding author.

\bibliographystyle{IEEEtran}
\bibliography{zglu.bib}


\end{document}